\documentclass[pdflatex,sn-mathphys-num]{sn-jnl}

\usepackage[version=3]{mhchem} 

\makeatletter
\gdef\orcidlogo{\includegraphics[height=9pt]{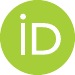}}
\makeatother

\makeatletter
\def\tablebodyfont{\reset@font\fontsize{10bp}{11bp}\selectfont}
\def\tablecolheadfont{\reset@font\fontsize{9bp}{11bp}\selectfont\bfseries\boldmath}
\def\tablecaptionfont{\reset@font\fontsize{9bp}{11bp}\selectfont}
\def\tablefootnotefont{\reset@font\fontsize{8.5bp}{10bp}\selectfont}

\makeatother

\usepackage{gensymb}
\usepackage{graphicx}
\usepackage{dcolumn}
\usepackage{bm}
\usepackage{siunitx}
\usepackage{makecell}
\usepackage{xcolor}
\usepackage{soul} 
\usepackage{ulem}
\usepackage{amssymb, amsmath, amsfonts}
\usepackage{nameref}
\usepackage{subcaption}
\usepackage{lineno}
\usepackage{caption}
\usepackage{longtable}

\usepackage{amsthm}
\usepackage{mathrsfs}%
\usepackage[title]{appendix}%
\usepackage{textcomp}%
\usepackage{manyfoot}%
\usepackage{booktabs}%
\usepackage{algorithm}%
\usepackage{algorithmicx}%
\usepackage{algpseudocode}%
\usepackage{listings}%

\setcitestyle{numbers,square,comma}

\usepackage{hyperref}
\usepackage[capitalize]{cleveref}

\AtBeginDocument{%

  \DeclareRobustCommand{\doi}[1]{\href{https://doi.org/#1}{\nolinkurl{#1}}}
  
  \DeclareRobustCommand\cite[1]{\unskip\hspace{1pt}\citep{#1}}
}

\newcommand{\etal}{{\it et al.}}
\newcommand{\ie}{{\it i.e.}, }
\DeclareSIUnit{\ramcm}{\per\centi\metre}
\DeclareSIUnit{\wtpercent}{wt\%}

\DeclareSIUnit\rpm{rpm}
\DeclareSIUnit\dyne{dyne}
\definecolor{pinegreen}{rgb}{0.0, 0.47, 0.44}

\begin{document}

\title[Direct Realization of Near-Ideal Carbyne in Ultrathin Boron Nitride Nanotubes]{Direct Realization of Near-Ideal Carbyne in Ultrathin Boron Nitride Nanotubes}


\author[1]{\fnm{Iryna} \sur{Ivanenko}\,\orcid{https://orcid.org/0000-0002-6885-3662}}\equalcont{These authors contributed equally to this work.}

\author[2,3]{\fnm{Pietro} \sur{Marabotti}\,\orcid{https://orcid.org/0000-0003-3451-845X}}\equalcont{These authors contributed equally to this work.}

\author[4,5]{\fnm{Yifan} \sur{Zhang}\,\orcid{https://orcid.org/0000-0003-1849-7484}}

\author[2,3]{\fnm{Getulio} \sur{Silva e Souza J\'{u}nior}\,\orcid{https://orcid.org/0009-0005-3159-845X}}

\author[2,3]{\fnm{Johannes M. A.} \sur{Lechner}}

\author[2,3]{\fnm{Pablo} \sur{Hern\'{a}ndez L\'{o}pez}\,\orcid{https://orcid.org/0000-0002-8064-2198}}

\author[1]{\fnm{Martin} \sur{Magg}\,\orcid{https://orcid.org/0009-0005-9825-0511}}

\author[1]{\fnm{Shivani} \sur{Shivaprakash}}

\author[4]{\fnm{Carlo Spartaco} \sur{Casari}\,\orcid{https://orcid.org/0000-0001-9144-6822}}

\author*[2,3]{\fnm{Sebastian} \sur{Heeg}\,\orcid{https://orcid.org/0000-0002-6485-3083}}\email{sebastian.heeg@physik.hu-berlin.de}

\author*[1]{\fnm{Benjamin S.} \sur{Flavel}\,\orcid{https://orcid.org/0000-0002-8213-8673}}\email{benjamin.flavel@kit.edu}

\affil[1]{\orgdiv{Institute of Nanotechnology}, \orgname{Karlsruhe Institute of Technology}, \orgaddress{\street{Kaiserstraße 12}, \city{Karlsruhe}, \postcode{76131}, \country{Germany}}}

\affil[2]{\orgdiv{Institut f\"ur Physik}, \orgname{Humboldt-Universit\"at zu Berlin}, \orgaddress{\street{Newtonstraße 15}, \city{Berlin}, \postcode{12489}, \country{Germany}}}

\affil[3]{\orgdiv{Center for the Science of Materials Berlin}, \orgname{Humboldt-Universit\"at zu Berlin}, \orgaddress{\street{Zum Großen Windkanal 2}, \city{Berlin}, \postcode{12489}, \country{Germany}}}

\affil[4]{\orgdiv{Dipartimento di Energia}, \orgname{Politecnico di Milano}, \orgaddress{\street{via R. Lambruschini 4}, \city{Milan}, \postcode{20156}, \country{Italy}}}

\affil[5]{\orgdiv{School of Engineering}, \orgname{Huzhou Normal University}, \orgaddress{\city{Huzhou}, \state{Zhejiang}, \postcode{313000}, \country{China}}}


\abstract{Carbyne, the sp-hybridized one-dimensional allotrope of carbon, is predicted to be the stiffest known material, with electronic and optical properties set by a single structural parameter, the bond length alternation. However, its intrinsic properties have never been measured: chains synthesized through molecular chemistry carry endgroup and finite-length perturbations that persist even in the longest molecules available, while chains grown inside carbon nanotubes strongly couple to the host, which renormalizes their vibrational frequency by up to \qty{110}{\ramcm} in a diameter-dependent manner. Here, we show that encapsulating and thermally converting hydrogen-capped polyynes inside ultrathin boron nitride nanotubes, structural analogues of carbon nanotubes but electrically insulating, yields carbyne chains in a near-ideal regime, where endgroup, finite length, and host-guest perturbations are reduced to secondary effects. Statistical Raman spectroscopy across 245 locations returns a vibrational frequency distribution an order of magnitude narrower than in carbon nanotubes, an anharmonicity consistent with the universal law for carbyne-like materials, and a bond length alternation matching correlated calculations for the free chain. No photoluminescence is detected, despite the transparent host, as expected for the dipole-forbidden emission of an unperturbed carbyne chain. Boron nitride nanotubes give experimental access to carbyne in its near-ideal form.}


\maketitle

\section*{Introduction}\label{sec:intro}
\begin{figure}[!ht]
    \centering    
    \includegraphics[]{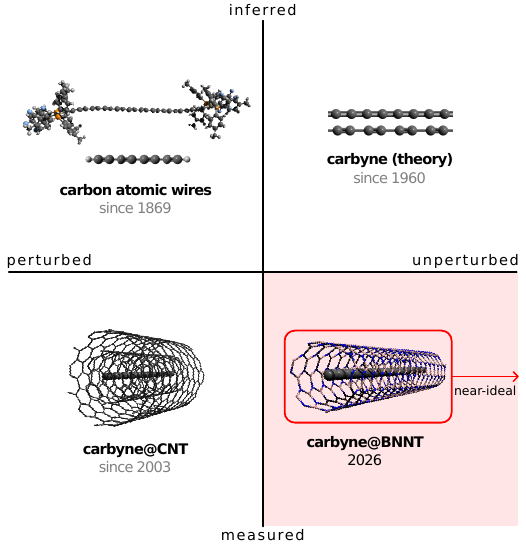}

    \caption{Carbyne realizations classified by their degree of perturbation from the unperturbed, ideal carbyne limit, and by how their infinite-chain properties are obtained, \ie inferred or directly measured: carbon atomic wires (inferred, perturbed)~\cite{glaser1869beitrage}, carbyne confined within CNTs (carbyne@CNT; measured, perturbed)~\cite{zhao2003carbon}, and carbyne confined within BNNTs (carbyne@BNNT, this work; measured, near-unperturbed). Dates indicate carbyne's theoretical definition~\cite{kudryavtsev1971carbyne} and the first reported synthesis of each realization.}
    \label{fig:fig1}   
\end{figure}

\noindent Carbyne, a linear chain of sp-hybridized carbon atoms, is the one-dimensional, one-atom-thick limit of carbon (\cref{fig:fig1})~\cite{casari2016carbon, bryce2021review}. It is expected to exhibit a range of extreme properties, including an exceptionally high Young's modulus, large Debye temperature, and a huge zero-point vibrational energy~\cite{liu2013carbyne, liu2015tunable, artyukhov2014mechanically}. Its structural, vibrational, and electronic properties are governed by a single structural parameter, the bond length alternation (BLA), which is defined as the average length difference between adjacent CC bonds and sets both the optical gap and the frequency of the Raman-active stretching mode (C mode)~\cite{milani2009connection, casari2016carbon}. This mode shows the highest resonance Raman scattering cross section reported to date, exceeding that of graphene and carbon nanotubes~\cite{tschannen2020raman}, and the strong electron-phonon coupling behind it makes carbyne an exceptionally sensitive probe of its local environment~\cite{tschannen2021anti}.

\noindent  Free-standing ideal carbyne has not been realized since it does not survive ambient conditions, where extended chains rapidly crosslink, oxidize, or rearrange~\cite{casari2004chemical, cataldo2006stability, heymann2005thermolysis}. Lacking the real material, several theoretical studies have been carried out, but their outputs do not converge on the properties of the ideal chain. For instance, estimations of the free chain place the BLA from \qty{\approx 0.0844}{\angstrom} to \qty{\approx 0.1348}{\angstrom}, while the corresponding C mode frequencies span more than \qty{200}{\ramcm}~\cite{milani2008first, milani2008carbynes, ramberger2021new, romanin2021dominant, mostaani2016quasiparticle}. Theory alone cannot deliver the ideal limit, and an experimental realization approaching it is required. However, every existing carbyne-like material relies on chemical stabilization or nanoscale confinement, and each introduces perturbations that mask intrinsic properties of the ideal chain~\cite{bryce2021review, casari2016carbon, gao2022advances, arora2023monodisperse, zhao2003carbon, shi2016confined, maruyama2025highly}.

\noindent Carbon atomic wires (CAWs) are finite molecules whose lengths and terminations determine the wire BLA (\cref{fig:fig1})~\cite{bryce2021review, casari2016carbon, gao2022advances, arora2023monodisperse, gao2020loss}. Accessing the infinite chain limit requires extrapolating finite-length data, but endgroup perturbations persist even in the longest chains synthesized to date. The extrapolated asymptotes for the longitudinal stretching mode do not converge to a single value, ranging from \qtyrange[range-units=single]{1790}{1908}{\ramcm} depending on the terminations, a spread large enough to make this approach an unreliable reference for ideal carbyne~\cite{gao2022advances, arora2023monodisperse}.

\noindent Nanoscale confinement within carbon nanotubes (CNTs) enables the growth of extended chains free from finite-size effects, either during CNT growth or via bottom-up conversion of encapsulated molecular precursors (\cref{fig:fig1})~\cite{zhao2003carbon, shi2016confined, zhang2025low, schuster2025quantifying, zhao2011growth, chang2021smallest, tang2024encapsulation, maruyama2025highly, shi2021photothermal, freytag2025systematic, chen2025universal}. However, strong host-guest interactions, including van der Waals coupling, dielectric screening, and electron hybridization, renormalize the BLA in a diameter-dependent manner, producing a broad distribution of C mode frequencies between \qty{\approx 1760}{\ramcm} and \qty{1870}{\ramcm} and optical gaps from \qty{\approx 1.64}{\electronvolt} to \qty{2.32}{\electronvolt}, which does not converge with tube diameter~\cite{heeg2018carbon, zhang2025low, martinati2022electronic, shi2017electronic, lechner2025universal, parth2025anharmonic, shi2016confined, lv2022one, shi2021toward}.

\noindent An ideal host for carbyne should combine nanoscale confinement with weak coupling. Boron nitride nanotubes (BNNTs) are structural analogues of CNTs, but their wide bandgap (\qtyrange[range-units=single, range-phrase=--]{\approx 5.5}{6}{\electronvolt}) and absence of a delocalized $\pi$-electron conjugated system make them electronically inert nanocontainers that stabilize encapsulated species without renormalizing their intrinsic properties~\cite{golberg2010boron, wang2007static, allard2020confinement, juergensen2025collective}. Despite these favorable characteristics, neither carbyne nor any carbon atomic wires have been encapsulated in BNNTs to date. The systems confined in BNNTs so far are $\pi$-electron conjugated dye molecules~\cite{juergensen2025collective, allard2020confinement} and graphene nanoribbons obtained by thermal conversion of encapsulated coronene~\cite{barzegar2016synthesis}, accommodated in tubes with inner diameters spanning \qtyrange[range-units=single, range-phrase=--]{0.7}{4.5}{\nano\meter}. Isolating BNNTs in the ultrathin diameter range associated with extended chain formation in CNTs (\qtyrange[range-units=single, range-phrase=--]{\approx 0.6}{1.3}{\nano\meter}) has remained a long-standing difficulty, and it is the reason why carbyne has so far been out of reach in a BN host~\cite{shi2016confined, zhang2025low, chen2025universal, kim2017scalable, marincel2019scalable, ko2023scalable, kang2026enhanced, lee2020purification}.

\noindent Here, we realize near-ideal carbyne, which we define as a regime in which size-confinement, endgroup perturbations, and host-guest renormalization are reduced to secondary effects, such that the BLA and every observable it governs approach the values of the infinite, unperturbed chain (\cref{fig:fig1}). The synthesis follows a bottom-up approach: hydrogen-capped polyynes are encapsulated within ultrathin BNNTs, isolated from commercially available material, and then converted to extended chains by vacuum annealing. Statistical Raman analysis across 245 locations returns a narrow distribution of C mode frequencies centered at \qty{\approx 1875}{\ramcm}, shifted above the range accessible to CNT-based systems and an order of magnitude narrower than any carbyne distribution reported to date~\cite{zhang2025low, heeg2018carbon, lechner2022raman}. The near-ideal regime is established on two observables: the vibrational anharmonicity of the chains follows the universal law for carbyne-like materials~\cite{lechner2025universal}, and the BLA extracted from the C mode matches correlated calculations for the free chain. No photoluminescence is detected in the transparent host, as expected for the symmetry-forbidden emission of the ideal chain. Together, these observables establish carbyne chains in boron nitride nanotubes as the closest realization of ideal carbyne reported to date.

\section*{Results and Discussion}\label{sec:results}
\subsection*{Isolation of ultrathin boron nitride nanotubes and synthesis of carbyne chains}
\noindent To synthesize carbyne in boron nitride nanotubes (carbyne@BNNTs), we follow a bottom-up approach already employed for CNTs, \ie encapsulating a carbon-based molecular precursor within the nanotube core and converting it into extended chains through annealing~\cite{chen2025universal, zhao2011growth, feng2024robust, zhang2025low, maruyama2025highly}. Even though the diameter dependence of precursor encapsulation and extended chain stabilization has not been established for BNNTs, we target ultrathin BNNTs (\qtyrange[range-units=single, range-phrase=--]{0.6}{1.3}{\nano\meter}), building on the diameter range supporting successful carbyne formation within CNTs~\cite{chen2025universal, shi2021toward, freytag2025systematic, schuster2025quantifying, shi2016confined, zhang2025low, chang2021smallest, tang2024encapsulation, shi2021photothermal}. We isolate them from the ``puff ball" parent material by tip ultrasonication in aqueous surfactant followed by density gradient ultracentrifugation (DGU, see~\nameref{sec:methods}, \cref{fig:fig2a}, and \cref{fig:SI_centr_tubes} in the Supplementary Information, SI). DGU and atomic force microscopy (AFM) measurements resolve two distinct BNNT populations (\cref{fig:fig2b} and \cref{fig:SI_afm} in the SI): Band~1, enriched in tubes with diameter of \qtyrange[range-phrase=--, range-units=single]{0.5}{1.5}{\nano\meter}, and Band~2, dominated by larger tubes with diameter of \qtyrange[range-phrase=--, range-units=single]{1.5}{2.5}{\nano\meter}. Based on the desired diameter distribution, we select BNNTs in Band~1 as the host material for carbyne growth.

\begin{figure}[!ht]
    \centering    
    \includegraphics[]{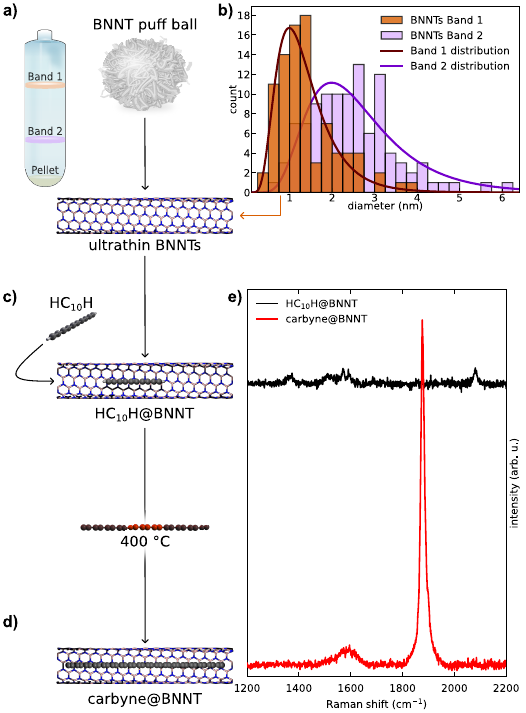}

    \caption{a) The parent ``puff ball" containing BNNTs is centrifuged and, after stratification, forms distinct bands of BNNTs (here in false colors). b) Distribution of diameters (bin width = \qty{0.25}{\nano\meter}) of Band~1 (orange), containing ultrathin BNNTs, and Band~2 (light violet). Lognormal fitted curves are reported for both Bands (\cref{tab:SI_lognorm_fit} in the SI). Molecular sketches of c) the encapsulation of HC$_{10}$H molecules into a BN nanotube and wires merging through vacuum annealing at \qty{400}{\celsius} (\nameref{sec:methods}), providing d) carbyne encapsulated within BNNTs. e) Raman spectra of the HC$_{10}$H encapsulated in BNNTs (black) and carbyne@BNNT (red). All these spectra have been collected using a \qty{532}{\nano\meter} excitation (see~\nameref{sec:methods}).}
    \label{fig:fig2}
    
    \phantomsubcaption{\label{fig:fig2a}}
    \phantomsubcaption{\label{fig:fig2b}}
    \phantomsubcaption{\label{fig:fig2c}}
    \phantomsubcaption{\label{fig:fig2d}}
    \phantomsubcaption{\label{fig:fig2e}}
    
\end{figure}

\noindent We filter the BNNTs in Band~1 and deposit them onto sapphire substrates by membrane transfer (see~\nameref{sec:methods}). The presence of the BNNT film and its high degree of crystallinity are confirmed by UV-Vis absorption spectroscopy, which exhibits a pronounced absorption peak at \qty{212}{\nano\meter} (\qty{5.85}{\electronvolt}, see~\cref{fig:SI_uvvis_a} in the SI), consistent with the optical transitions of BNNTs~\cite{lauret2005optical}. A representative Raman spectrum of this film is shown in \cref{fig:SI_growth_a} in the SI.

\noindent We encapsulate the carbyne molecular precursor immersing the BNNT film in a solution of size-selected hydrogen-capped polyynes, \ie HC$_{10}$H in cyclohexane, and keep it at room temperature for \qty{24}{\hour} (\cref{fig:fig2c}; see~\nameref{sec:methods} and~\cref{fig:SI_uvvis_b} in the SI for HC$_{10}$H pristine absorption spectrum). This bottom-up approach relies on the polyyne-to-carbyne conversion process validated in CNTs~\cite{zhao2011growth, chang2021smallest, tang2024encapsulation, maruyama2025highly}. The appearance of a new Raman peak at \qty{2082}{\ramcm} (black spectrum in~\cref{fig:fig2e}) reports the encapsulation of HC$_{10}$H inside BNNTs. This mode corresponds to the characteristic stretching mode of confined HC$_{10}$H and is redshifted by \qty{38}{\ramcm} compared to the isolated molecule in solution (\qty{2120}{\ramcm})~\cite{marabotti2022electron, tabata2006raman}. This redshift (\qty{38}{\ramcm}) is smaller than that occurring in any CNT hosting the same polyyne (\qtyrange[range-units=single, range-phrase=--]{49}{62}{\ramcm}), indicating that the perturbation induced by the BN host is smaller than that reported for CNTs, as discussed in more detail later and in \cref{sec:SI_growth} in the SI~\cite{nishide2006single, nishide2007raman, tang2024encapsulation, moura2009charge, moura2011dielectric}.

\noindent After annealing the HC$_{10}$H@BNNT film under high vacuum (\qty[retain-unity-mantissa=false]{< 1e-4}{\pascal}) at \qty{400}{\celsius} for \qty{8}{\hour} (\cref{fig:fig2c}; see~\nameref{sec:methods} and~\cref{sec:SI_growth} in the SI) the HC$_{10}$H Raman peaks largely disappear in most locations probed across the sample. New intense Raman signals emerge around \qtyrange[range-units=single, range-phrase=--]{1870}{1890}{\ramcm} (red spectrum in~\cref{fig:fig2e}), in the characteristic frequency range of the stretching mode (C mode) of confined carbyne~\cite{maruyama2025highly, zhang2025low, lechner2022raman, heeg2018carbon, shi2016confined}. This indicates efficient conversion of the encapsulated precursor into carbyne@BNNT (\cref{fig:fig2d}) by thermally activated HC$_{10}$H fusion and chain extension within the nanotube cavity, similar to CNT-based reactions that show comparable changes in the Raman spectra~\cite{chang2021smallest, tang2024encapsulation, maruyama2025highly}. This also explains the sp\textsuperscript{2}-like carbon contributions, consistent with amorphous carbon or disordered graphene nanoribbons, observed in some sample regions, likely arising from precursor decomposition when chain stabilization is not favored (see \cref{sec:SI_growth} in the SI). Such compounds' formation by thermal annealing has been documented in other studies, though starting from different precursors~\cite{barzegar2016synthesis, cadena2023encapsulation}.

\noindent We observe a gradual laser-induced degradation of the carbyne@BNNT chains under continuous irradiation, manifested as a decrease in the C mode intensity (see \cref{sec:SI_degr} in the SI). This effect was carefully accounted for in all measurements presented hereafter, keeping the laser power at the minimum possible level and reducing the exposure time to limit irradiation-induced damage.

\subsection*{Vibrational, structural, and electronic properties of near-ideal carbyne}
\begin{figure}[!ht]
    \centering    
    \includegraphics[]{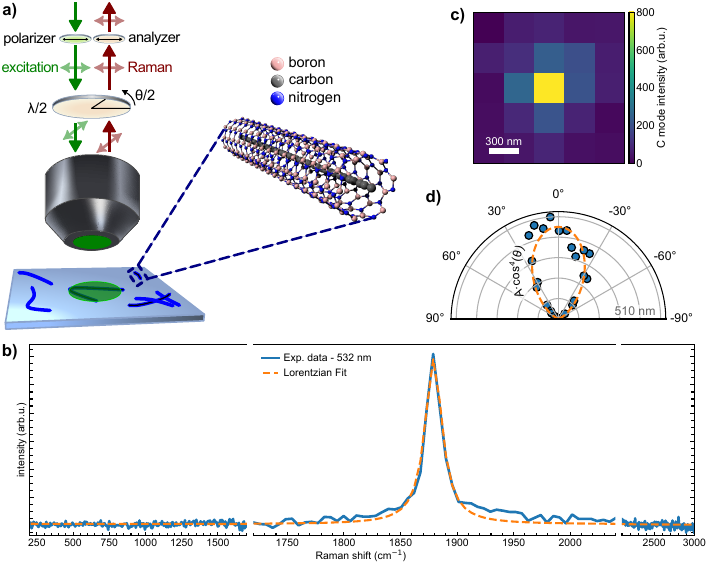}

    \caption{a) Illustration of the Raman setup used to map the carbyne@BNNT film, including the polarization optics used for polarization analysis. b) Raman spectrum of an isolated carbyne@BNNT chain (light blue line) collected using a \qty{532}{\nano\meter} laser showing a single C mode with a Lorentzian lineshape (orange dashed line), centered at \qty{1879 \pm 1}{\ramcm} with a full width at half maximum (FWHM) of \qty{18.3 \pm 0.4}{\ramcm}. The background was subtracted using a spectrum of the bare substrate measured during the same Raman map. c) Confocal Raman map showing the C mode intensity (step size = \qty{300}{\nano\meter}). d) Polarization-dependent C mode scattering intensity (blue circles) for a similar carbyne@BNNT chain collected with a \qty{510}{\nano\meter} laser (see~\nameref{sec:methods}). The dashed orange line indicates the $\cos^4$ periodicity.}
    \label{fig:fig3}

    \phantomsubcaption{\label{fig:fig3a}}
    \phantomsubcaption{\label{fig:fig3b}}
    \phantomsubcaption{\label{fig:fig3c}}
    \phantomsubcaption{\label{fig:fig3d}}
    
\end{figure}

\noindent Scanning across the carbyne@BNNT film through Raman mapping (\cref{fig:fig3a}), we find regions showing multiple overlapping C mode signals, resembling several carbyne@BNNT systems clustered together (\cref{fig:SI_bigmap} in the SI), and locations that show the characteristic Raman signatures of an isolated, single carbyne chain (\cref{fig:fig3c})~\cite{shi2016confined,heeg2018carbon,tschannen2020raman,lechner2025universal, lechner2022raman, heeg2018raman}. A representative Raman spectrum of such an isolated system (\cref{fig:fig3b}) shows a single C mode with a Lorentzian lineshape, collected from an isolated bright spot in the center of the confocal Raman map (\cref{fig:fig3c}). For a comparable chain, \cref{fig:fig3d} shows the integrated C mode Raman scattering $I_C$($\theta$) measured in parallel polarization, with a rotating half-wave plate setting the angle $\theta$ of the linear polarization at the sample and a fixed analyzer parallel to the incident polarizer (\cref{fig:fig3a}, see \nameref{sec:methods}). $I_C$($\theta$) follows a clear $\cos^4$ angular dependence with maximal intensity at $\theta=0$. This is the hallmark feature of a fully symmetric Raman mode in a highly anisotropic one-dimensional system, which identifies $\theta = 0$ as the chain axis~\cite{heeg2018raman}. This pronounced anisotropy, together with the spectral and spatial isolation of the C mode in both locations, indicates that these signals arise from individual, axially aligned carbyne chains.

\begin{figure}[!ht]
    \includegraphics[width=\textwidth]{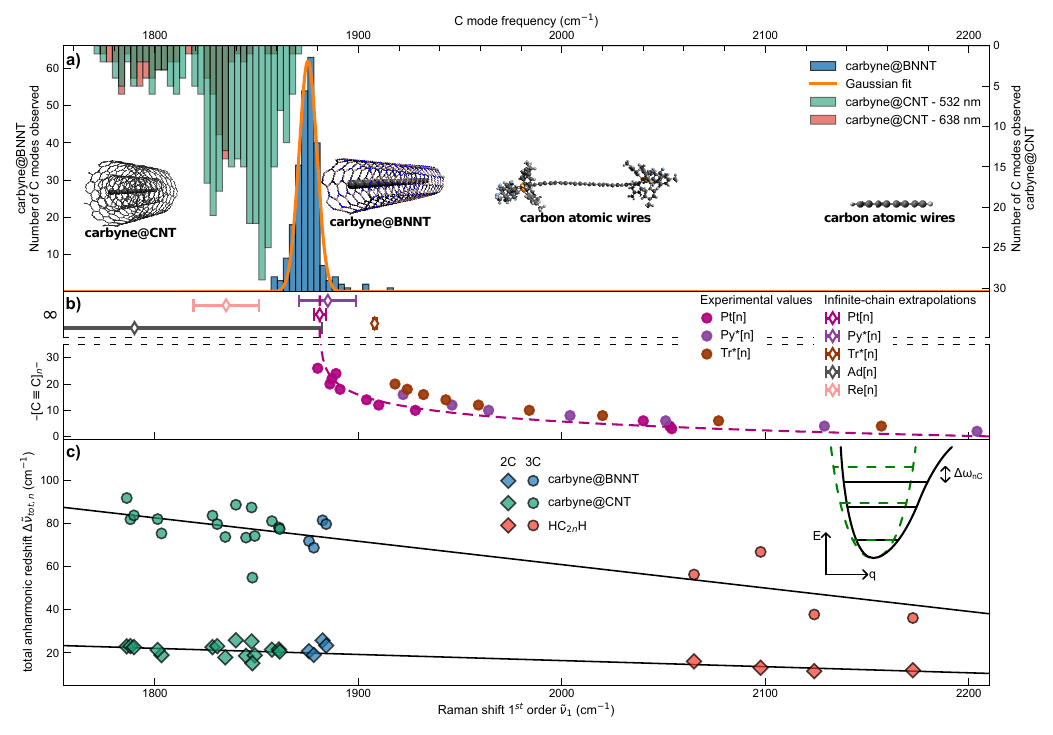}
    
    \caption{a) Distribution of C mode frequencies for carbyne@BNNT (blue bars). A distribution for carbyne@CNT is reported, collected across multiple regions, samples, and with two different excitation wavelengths (\qty{532}{\nano\meter} for light green bars, \qty{638}{\nano\meter} for light red bars). The common bin width is \qty{3}{\ramcm}. The orange curve represents the Gaussian fit of the carbyne@BNNT C mode distribution. b) Experimental stretching mode frequencies of Pt[n] (dark magenta circles)~\cite{arora2023monodisperse}, Py*[n] (violet circles)~\cite{gao2020loss}, and Tr*[n] (brown circles)~\cite{gao2020loss}, where $n$ corresponds to the number of acetylene units ($-[C\equiv C]_n-$). Infinite-chain extrapolated values for the previous molecular systems are reported as well, including also Ad[n] and Re[n] series (open diamonds with error bars coming from the extrapolation model)~\cite{gao2020loss}. c) Total anharmonic redshift ($\Delta\tilde{\nu}_{tot,n}$) as a function of the fundamental C mode frequency $\tilde{\nu}_1$ for carbyne@CNT (green)~\cite{lechner2025universal}, hydrogen-capped polyynes (HC$_{2n}$H) in solution (orange)~\cite{lechner2025universal}, and carbyne@BNNT (light blue), up to the third-order overtones. The universal anharmonic redshifts for 2C and 3C (black lines) are taken from Ref.~\cite{lechner2025universal}. In the top right, schematic harmonic and anharmonic potentials are shown.}
    \label{fig:fig4}
    
    \phantomsubcaption{\label{fig:fig4a}}
    \phantomsubcaption{\label{fig:fig4b}}
    \phantomsubcaption{\label{fig:fig4c}}
\end{figure}

\noindent Across 245 locations, the C mode of carbyne@BNNT follows a remarkably narrow Gaussian distribution centered at \qty[uncertainty-mode = separate, separate-uncertainty-units = single]{1875.2 \pm 0.1}{\ramcm} with a FWHM of \qty[uncertainty-mode = separate, separate-uncertainty-units = single]{10.3 \pm 0.2}{\ramcm} (\cref{fig:fig4a}). This is a clear sign that the C mode has experimentally converged to $\langle \omega_C \rangle_{\mathrm{BNNT}}$ = \qty{1875}{\ramcm} that is intrinsic to carbyne@BNNT and any residual environmental perturbation, \textit{e.g.}, screening by the confining nanotube, is reduced to a secondary perturbation. Such convergence has  been reached neither by carbon atomic wires nor carbyne@CNT, nor by any other carbyne-like material. Measured under the same excitation wavelength, the carbyne@CNT distribution spans \qtyrange[range-phrase=--, range-units=single]{1760}{1870}{\ramcm} with a host diameter window of \qtyrange[range-units=single, range-phrase=--]{0.6}{1.3}{\nano\meter}, an 
order of magnitude broader than carbyne@BNNT and regardless of which subset of the BNNT ensemble hosts the chains (\cref{fig:fig4a}).

\noindent The position of $\langle \omega_C \rangle_{\mathrm{BNNT}}$ is bracketed by the two experimental carbyne realizations and falls within the range spanned by calculations for ideal carbyne, which scatter by more than \qty{200}{\ramcm} depending on the level of theory (see \cref{fig:SI_distr_theo_extrap} in the SI)~\cite{shi2016confined, heeg2018carbon, maruyama2025highly, chen2025universal, zhang2025low, arora2023monodisperse, gao2020loss, milani2008carbynes, romanin2021dominant}. From below, carbyne@CNT blueshifts with increasing diameter without reaching the limit of vanishing host-guest interactions and its upper edge bounds the unperturbed chain from below (\cref{fig:fig4a})~\cite{shi2016confined, heeg2018carbon, martinati2022electronic, lechner2025universal, maruyama2025highly, chen2025universal, zhang2025low}. From above, the longest carbon atomic wires synthesized to date reach \qtyrange[range-phrase=--, range-units=single]{1881}{1890}{\ramcm} depending on the terminations and are still redshifting with increasing length (\cref{fig:fig4b})~\cite{arora2023monodisperse, gao2020loss}. Each series converges below its own longest member, and the lowest of them bounds the unperturbed chain from above. These two perturbations act in opposite directions, leaving the near-ideal limit constrained to the \qtyrange[range-phrase=--, range-units=single]{1870}{1881}{\ramcm} window, precisely where the carbyne@BNNT distribution is centered (see \cref{fig:fig4a,fig:fig4b}).

\noindent The extrapolated asymptotes provide no defined reference. Even though individual values fall close to, or within, the carbyne@BNNT distribution, this proximity cannot be treated as an independent validation, since endgroup perturbations displace the same asymptote from \qtyrange[range-units=single]{1790}{1908}{\ramcm}~\cite{arora2023monodisperse, gao2022advances}. The absence of a unique extrapolated limit is precisely why a direct measurement is required. 

\noindent The near-ideal character of carbyne@BNNT also manifests in the anharmonic potential landscape of the C mode. Raman spectroscopy extracts vibrational anharmonicity through the deviation of overtone frequency from the harmonic predictions. Therefore, the total anharmonic redshift is $\Delta\tilde{\nu}_{tot,n} = n\cdot\tilde{\nu}_1 - \tilde{\nu}_n$, where $\tilde{\nu}_{1}$ is the fundamental C mode frequency and $\tilde{\nu}_{n}$ that of the $n^{th}$ overtone. Resolving overtones requires chains bright enough to detect at least the second and third order, which restricts this analysis to four carbyne@BNNT chains (\cref{fig:fig4c} and \cref{fig:SI_overtones} and \cref{tab:SI_anh} in the SI). All four fall on the universal anharmonicity trend for carbyne-like materials established from carbyne@CNT and CAWs~\cite{lechner2025universal}, with a mean offset of \qty[uncertainty-mode = separate, separate-uncertainty-units = single]{2.4+-1.6}{\ramcm} for the 2C redshift and \qty[uncertainty-mode = separate, separate-uncertainty-units = single]{1.5+-3.8}{\ramcm} for 3C, and with a dispersion matching that of the systems defining the trend (see \cref{sec:SI_anh} in the SI). This result tests whether the host perturbs the vibrational potential beyond a rigid shift of the BLA as shown by strong anharmonic host-guest interactions in carbyne@CNT~\cite{parth2025anharmonic}. The BNNT host does not produce such a displacement. Therefore, the anharmonicity of the encapsulated chain falls in between the CAW and carbyne@CNT regimes and follows from its C mode frequency alone, the Raman observable of the BLA. 

\noindent The vibrational fingerprint of carbyne@BNNT translates directly into structure. Through the H\"{u}ckel model chain of Ref.~\cite{milani2009connection} and the empirical correlation between C mode frequency and optical gap established for carbyne@CNT~\cite{martinati2022electronic}, $\langle \omega_C \rangle_{\mathrm{BNNT}}$ corresponds to an optical gap of \qty{\approx 2.35}{\electronvolt} and a BLA of \qty{\approx 0.129}{\angstrom} (see \cref{sec:SI_bla} in the SI). This sits at the lower edge of the BLA values computed for a free chain by different computational methods (see \cref{tab:SI_theory} in the SI)~\cite{ramberger2021new, mostaani2016quasiparticle, romanin2021dominant}. Applied to the carbyne@CNT frequency range, the same model returns \qtyrange[range-units=single, range-phrase=--]{0.094}{0.127}{\angstrom}, whose lower end matches the same calculation repeated for a chain inside a nanotube (\qtyrange[range-units=single, range-phrase=--]{0.091}{0.098}{\angstrom})~\cite{ramberger2021new, romanin2021dominant}. Two carbyne realizations, analyzed with one model, provide BLA values that independent calculations assign to free and confined chains.

\noindent The host modulation of the BLA follows from the C mode distributions, resulting in a distribution an order of magnitude narrower in carbyne@BNNT (\qty{0.003}{\angstrom}) than in carbyne@CNT (\qty{0.033}{\angstrom}). Extrapolations from molecular series provide no comparable constraint, spanning \qtyrange[range-units=single, range-phrase=--]{0.092}{0.145}{\angstrom} depending on the termination~\cite{arora2023monodisperse, gao2022advances}, a range wider than the host-induced modulation and without a converged asymptote. Although the BLA absolute values reported here are not independent observables but a structural re-parametrization of the C mode frequency within the model approximations (\cref{sec:SI_bla_approx} in the SI), the comparisons above remain valid because the identical model chain is applied to every system.


\noindent The electronic response of carbyne@BNNT is governed by the same BLA and chain potential. In an ideal, centrosymmetric chain, the $\mathrm{D_{\infty h}}$ point group symmetry places two dark excited states below a one-photon bright state, so that the radiative decay to the ground state is dipole-forbidden and no emission is expected (\cref{fig:pl_spectra} in the SI). We probe a region with several carbyne@BNNT chains clustered together, whose C mode centered at \qty{\approx 1881}{\ramcm} corresponds to an optical gap of \qty{\approx 2.39}{\electronvolt} (\cref{fig:pl_spectra} and \cref{sec:pl} in the SI)~\cite{martinati2022electronic}. Exciting at \qty{2.43}{\electronvolt}, the intense resonance Raman C mode and its overtones establish that the electronic transition occurs and the excited state is populated, yet no emission is detected down to \qty{8}{\percent} of the C mode intensity between \qty{2.42}{\electronvolt} and \qty{1.60}{\electronvolt}, a window containing the first two vibronic replicas (\qty{2.16}{\electronvolt} and \qty{1.92}{\electronvolt}) and the dark state transitions extrapolated for the infinite chain in Ref.~\cite{zirzlmeier2020optical}. This differs from carbyne@CNT, where the host itself prevents emission through efficient non-radiative channels: the BNNTs are wide-bandgap insulators with no low-energy quenching channels, and emission is observed from other encapsulated species~\cite{allard2020confinement, juergensen2025collective}. Therefore, the absence of emission in a transparent host is consistent with the selection rule of the unperturbed chain, although other mechanisms remain possible, such as non-radiative decay through chain or host defects and a quantum yield below our detection sensitivity at room temperature. Low-temperature and pump-probe measurements on individual chains are required to settle this point.


\section*{Conclusions}\label{sec:conclusions}

\noindent Carbyne@BNNT is the nearest experimental realization of ideal carbyne achieved to date. Size-confinement and termination effects, which persist even in the longest carbon atomic wires and leave their infinite-chain extrapolations unconverged~\cite{arora2023monodisperse, gao2022advances}, are reduced to secondary effects. The host-guest interactions that in carbyne@CNT parametrize chain properties with the tube diameter~\cite{heeg2018carbon, zhang2025low} are reduced by the electronically inert BN host to secondary effects. The C mode frequency and the narrowness of its distribution, the vibrational anharmonicity, and the extracted BLA converge to our definition of near-ideal carbyne, with the absence of PL consistent with this picture. This provides experimental access to sp-hybridized carbon in its near-ideal form.

\backmatter





\section*{Acknowledgements}
B.F., I.I., M.M., and S.S. acknowledge funding from the Deutsche Forschungsgemeinschaft (DFG) under research grants 834/13-1. Y.Z. acknowledges Horizon Europe for the Marie Sk{\l}odowska-Curie Fellowship (grant no. 101065920 — SCCAMC). P.M., G.S.S.J., J.M.A.L., P.H.L., and S.H. acknowledge funding from the Deutsche Forschungsgemeinschaft (DFG, German Research Foundation) under the Emmy Noether Initiative (Project-ID 433878606). P.M. acknowledges the financial support of the Einstein International Postdoctoral Fellows program (IPF-2022-727).

\section*{Author Contributions}
I.I., P.M., S.H., and B.F. conceived the study and designed the experiments. I.I. and B.F. designed the route to ultrathin BNNTs. Y.Z. synthesized and purified the molecular precursor under the supervision of C.S.C. I.I. prepared and isolated the ultrathin BNNTs, performed AFM and UV-Vis to check their diameters and film quality, carried out the encapsulation and the thermal conversion, together with the preliminary Raman screening of filling and conversion, with support from M.M. and S.S. P.M. performed the Raman and photoluminescence characterization, including the polarization-resolved measurements, with support from G.S.S.J. and J.M.A.L. G.S.S.J. and P.M. implemented the control software for the tunable laser source, while P.H.L. and J.M.A.L. built the corresponding optical setup. P.M. developed the models and performed the data analysis, with input from S.H., G.S.S.J. and J.M.A.L. P.M. prepared the final figures. I.I., P.M., S.H. and B.F. wrote the original draft, while all the authors contributed to the review and editing. S.H., B.F. and C.S.C. supervised the project and acquired funding. All authors discussed the results and contributed to the manuscript. S.H. and B.F. jointly supervised this work.

\section*{Experimental Methods}\label{sec:methods}
\subsection*{BNNT Film Preparation}
\noindent \qty{40}{\milli\gram} of boron nitride nanotubes (BNNT LLC, Batch \#Y8B01240729D), supplied as a ``puff ball"  containing predominantly multi-walled BNNTs with \qtyrange[range-phrase=--, range-units=single]{2}{3} layers, outer diameters of \qtyrange[range-phrase=--, range-units=single]{\approx 4}{5}{\nano\meter}, lengths of \qtyrange[range-phrase=--, range-units=single]{\approx 100}{200}{\micro\meter}, and nominally \qty{5}{\wtpercent} of single-walled BNNTs according to the supplier, were dispersed in \qty[per-mode = symbol]{20}{\gram\per\liter} (2~\%) sodium deoxycholate (DOC, PanReacAppliChem) by tip ultrasonication (Ultrasonic Werfr.,~\qty{36}{\kilo\hertz}) at~\qty{32}{\watt}~(approximately \qtyrange[range-units=single, range-phrase={~--~}, per-mode=symbol]{0.8}{0.9}{\watt\per\milli\liter}) for \qty{2.5}{\hour} in an ice bath. The resulting dispersion was centrifuged at 48400$\times g$ to pellet aggregates and large bundles, and the supernatant was collected for density gradient ultracentrifugation (DGU).

\noindent DGU was performed using \qty{30}{\milli\liter} centrifuge tubes (Beckman Coulter, Optiseal, Ref. 362183) in a VTi 50 rotor. \qtyrange[range-units=single, range-phrase=--]{8}{10}{\milli\liter} of the supernatant BNNT dispersion were layered on top of \qty{28}{\milli\liter} of 25~\% iodixanol (OptiPrep Density Gradient Medium) containing \qty[per-mode = symbol]{10}{\gram\per\liter} (1~\%) DOC and centrifuged at 242000$\times g$ for \qty{2}{\hour} and \qty{45}{\minute}. Additional experiments were carried out to optimize the centrifugation step (see~\cref{sec:SI_centr} in the SI). Following centrifugation, \qty{1}{\milli\liter} fractions were extracted sequentially from the top to the bottom of the centrifuge tube, yielding 25 fractions per run.

\noindent Individual fractions, or combinations of equivalent fractions, were concentrated and desalted by ultrafiltration through a VivaSpin 500 filter (Cytiva Sweden AB) at 10300$\times g$ at room temperature, processing \qty{0.1}{\milli\liter} per \qty{20}{\minute} cycle until the iodixanol was fully removed. The concentrated BNNT suspensions were characterized by UV-Vis absorption spectroscopy and AFM following deposition onto silicon substrates (\cref{fig:SI_uvvis_a,fig:SI_afm} in the SI). AFM revealed two distinct nanotube populations: Band~1 (fractions 8--10), containing tubes with apparent diameters of \qtyrange[range-phrase=--, range-units=single]{0.5}{1.5}{\nano\meter}, and Band~2 (fractions 21--23), dominated by larger tubes with apparent diameters of \qtyrange[range-phrase=--, range-units=single]{1.5}{2.5}{\nano\meter}. These diameters also constrain the number of walls. Considering a BN interlayer spacing of \qty{0.33}{\nano\meter}~\cite{golberg2010boron}, a double-walled tube requires an outer diameter above \qty{1.26}{\nano\meter} to host a chain. Therefore, we can assume that Band~1 mostly consists of single-walled tubes and, at the upper end of its range, double-walled ones, even though the full morphological characterization of BNNT tubes falls beyond the scope of this work.

\noindent Band~1 was selected for carbyne growth. For film preparation, the concentrated Band~1 suspension was filtered through a mixed cellulose ester membrane (MCE, HAWP, Millipore, pore size \qty{0.45}{\micro\meter}, area \qty{12.57}{\square\cm}) and washed repeatedly with deionized water until the filter ran clear, and a BNNT film was obtained on the membrane surface. The film was then transferred to a standard sapphire substrate (Ossila, S2005A1, \qtyproduct{20 x 15}{\mm}, \qty{1.1}{\mm} thick) by placing the membrane face down onto the substrate, wetting the back surface with deionized water, and compressing it with a PTFE coverslip during drying at \qty{70}{\celsius} for \qty{30}{\minute} to promote adhesion. The MCE membrane was subsequently dissolved by immersion in multiple acetone baths over \qty{1}{\hour}, leaving a thin, freestanding BNNT film on the sapphire surface~\cite{tune2013single}.

\subsection*{Polyyne Precursor Preparation}
\noindent Polyynes of varying chain lengths were prepared according to a modified method based on that reported in Ref.~\cite{cataldo2005method}. \qty{12}{\milli\liter} of ethanol (96\%, Carlo Erba Reagents), \qty{4}{\milli\liter} of deionized water, and \qty{10}{\milli\liter} of cyclohexane ($\geq$99.5\%, Sigma-Aldrich) were added sequentially. Subsequently, \qty{0.45}{\gram} of ammonium chloride ($\geq$99.5\%, Sigma-Aldrich), \qty{0.30}{\gram} of copper(I) chloride ($\geq$99\%, Sigma-Aldrich), and \qty{0.48}{\gram} of copper(II) chloride dihydrate ($\geq$97\%, Sigma-Aldrich) were introduced into a conical flask equipped with a magnetic stir bar. The resulting mixture was stirred to form a uniform suspension. With continuous stirring, \qty{0.4}{\gram} of calcium carbide ($\geq$75\%, pellets of \qty{\approx 2}{\milli\meter} size, Sigma-Aldrich) was added slowly. A slight evolution of acetylene bubbles was observed, and the solution turned reddish-brown. The reaction mixture was then stirred at ambient temperature for \qty{30}{\minute}. Subsequently, \qty{4}{\milli\liter} of dilute hydrochloric acid (ca. 20\%, prepared from 37\% HCl, Supelco) was added slowly to the reaction mixture. After the addition, stirring was stopped, and the solution was allowed to separate into two layers naturally: the upper layer, a yellow-colored cyclohexane solution containing the synthesized polyynes, and the lower layer, consisting of the aqueous phase along with solid precipitates. The obtained cyclohexane solution of polyynes was then passed through a silica gel column (particle size \qtyrange[range-phrase=--, range-units=single]{0.063}{0.200}{\milli\meter}, Labkem) to eliminate byproduct impurities.

\noindent Following the removal of byproduct impurities, the resulting cyclohexane solution containing the polyyne mixture was concentrated and solvent-exchanged into acetonitrile for high-performance liquid chromatography (HPLC) separation. Specifically, the cyclohexane solution was added dropwise onto an acetonitrile phase (HPLC grade, Carlo Erba Reagents). Upon evaporation of the cyclohexane, the polyynes dissolved into the acetonitrile, yielding \qty{2}{\milli\liter} of a concentrated polyyne solution in acetonitrile. Separation of individual polyyne chain lengths was then carried out using a Shimadzu HPLC system equipped with a Phenomenex Luna C18 column (\qtyproduct{250 x 10}{\milli\meter}, \qty{5}{\micro\meter} particles, C18 alkyl functionalization). The mobile phase gradient was programmed such that the acetonitrile content increased linearly from 65\% to 95\% over \qty{15}{\minute}, at a flow rate of \qty[per-mode = symbol]{3.7}{\milli\liter\per\minute}. The injection volume was \qty{500}{\micro\liter}. Chromatograms were recorded by coupling the HPLC system with a diode-array UV–Vis detector, enabling unambiguous assignment of specific polyyne chain lengths based on their characteristic retention times and absorbance spectra. Under these conditions, the target molecule HC$_{10}$H was collected at its corresponding retention time (\qtyrange[range-phrase=--, range-units=single]{8.8}{10.2}{\minute}).\\
After the collection of the fraction corresponding to HC$_{10}$H, the obtained solution containing the target compound in a water/acetonitrile mixture was mixed with cyclohexane. The resulting biphasic system was then shaken to allow for the extraction of HC$_{10}$H into cyclohexane. Upon phase separation, the cyclohexane layer containing the extracted HC$_{10}$H was collected.

\subsection*{Encapsulation of HC$_{10}$H}
\noindent The BNNT film on sapphire was immersed in a cyclohexane solution of HPLC-purified \ce{HC10H} (concentration~\qty{\approx 1e-4}{\mole\per\liter}) and kept at room temperature for \qty{24}{\hour} in the dark. The substrate was then removed from the solution, rinsed with pure cyclohexane to remove molecules adsorbed on the film surface, and dried in air.

\subsection*{Polyyne-to-Carbyne Conversion}
\noindent A \ce{HC10H}@BNNT film on sapphire was placed into a horizontal high-temperature furnace (HTRH 18/100/600, Carbolite) equipped with a ceramic 799 (RCA/Alsint) tube $\varnothing$ 100 (outer) $\times$ $\varnothing$ 88 (inner) $\times$ \qty{1415}{\mm} (length) and Turbolab 80 (Leybold) turbomolecular pump system. The sample was heated following a temperature program in which the sample is ramped from room temperature to \qty{400}{\celsius} at \qty[per-mode = symbol]{0.1}{\celsius\per\minute}, held at \qty{400}{\celsius} for \qty{8}{\hour}, and then cooled back to room temperature at \qty[per-mode = symbol]{0.1}{\celsius\per\minute}.

\subsection*{Atomic Force Microscopy}
\noindent Topographies were recorded with a Dimension Icon (Bruker) with NSC 19 cantilevers ($\mu$masch) with a resonance frequency of \qty{65}{\kilo\hertz} and a force constant of \qty[per-mode = symbol]{0.5}{\newton\per\meter}. Imaging was performed in standard tapping mode in air, with a resolution of 1024~lines. All topographies were evaluated using open-source Gwyddion. For the diameter determination, \qty{30}{\micro\liter} of ten times diluted dispersion was drop-coated onto a silicon wafer and spun at \qty{2000}{\rpm} for \qty{3}{\minute}.

\subsection*{Spectroscopy}
\noindent UV--vis--NIR absorbance spectra were collected on a Cary 500 spectrometer from \qtyrange{180}{3200}{\nm} for all samples. Measurements were performed either in cuvettes with a \qty{1}{\mm} path length or in transmission through the BNNT film deposited on sapphire.

\noindent Raman spectra of~\cref{fig:fig2e} were acquired with a Renishaw InVia Reflex confocal Raman spectrometer using an excitation wavelength of \qty{532}{\nm} (\qty{2.33}{\electronvolt}) under a 20$\times$ objective. Additional Raman measurements at \qty{532}{\nm} were performed with a Horiba Xplora spectrometer using circularly polarized light and a power of \qty{0.16}{\milli\watt} measured under a 100$\times$ objective (NA 0.95). Raman maps were acquired on the Horiba Xplora in ambient conditions with \qty{0.16}{\milli\watt} of circularly polarized light under the 100$\times$ objective (NA 0.95). The map of \cref{fig:fig3c} was recorded with a step size of \qty{300}{\nano\meter} and \qty{10}{\second} per pixel. The 245 C mode frequencies of \cref{fig:fig4a} were collected from Raman maps, with size from \qtyproduct{20 x 20}{\micro\meter} to \qtyproduct{60 x 60}{\micro\meter}, a step size of \qty{500}{\nano\meter}, and \qtyrange[range-units=single, range-phrase=--]{3}{4}{\second} per pixel, such as the Raman map in \cref{fig:SI_bigmap_a}. These maps include both spatially isolated single chains and clustered C mode regions. A spectrum was retained whenever a C mode was distinguishable from the background. Frequencies were obtained by fitting one or more Lorentzian curves on a linear baseline, the same procedure used for the polarization-dependent data.

\noindent The background of \cref{fig:fig3b} was removed using substrate spectra recorded within the same map. Each was scaled by a multiplicative factor chosen to match the signal-free regions (\qtyrange[range-units=single, range-phrase=--]{100}{1000}{\ramcm}, \qtyrange[range-units=single, range-phrase=--]{2100}{3300}{\ramcm}, and \qtyrange[range-units=single, range-phrase=--]{4000}{5000}{\ramcm}) of the carbyne spectrum. The substrate spectrum minimizing the residuals in those regions was subtracted with its corresponding factor.

\noindent Raman measurements at \qty{510}{\nano\meter} (\cref{fig:fig3d} and~\cref{fig:pl_spectra} in the SI) were carried out using a C-Wave GTR tunable continuous-wave laser (H\"ubner Photonics), providing excitation in the \qtyrange[range-units=single, range-phrase=--]{500}{750}{\nano\meter} range. The beam was directed to the sample through a linear polarizer, a 10:90 (R:T) beamsplitter, and a half-wave plate mounted on a motorized rotational stage, and then focused by a 100$\times$ long-working-distance objective (NA 0.7). The backscattered Raman signal was collected with the same objective, transmitted through the half-wave plate and beamsplitter, filtered with a tunable longpass filter (cutoff \qty{\approx 150}{\ramcm}) or a \qty{550}{\nano\meter} longpass filter to suppress Rayleigh scattering, and analyzed using a Kymera 328i spectrograph equipped with an iVac CCD detector (Quantum Design). For polarization-dependent measurements (\cref{fig:fig3d}), the half-wave plate was rotated over the full angular range (see the scheme in~\cref{fig:fig3a}), and the backscattered Raman light was analyzed using a second polarizer parallel to the excitation polarizer placed before the filter. The angular origin in~\cref{fig:fig3d} is set to the maximum of the~$\cos^4$ fit, which defines the chain axis. For all other measurements at \qty{510}{\nano\meter}, the half-wave plate was adjusted to align the incident polarization to the longitudinal axis of the carbyne chains (\ie maximum of the~$\cos^4$ fit) found in a previous measurement.

\noindent Wavelength calibration was performed for all spectra using Neon lamp emission lines.


\bibliography{references}
\clearpage

\setcounter{section}{0}
\setcounter{figure}{0}
\setcounter{table}{0}
\setcounter{equation}{0}
\renewcommand{\thesection}{S.\arabic{section}}
\renewcommand{\thefigure}{S.\arabic{figure}}
\renewcommand{\thetable}{S.\arabic{table}}
\renewcommand{\theequation}{S.\arabic{equation}}
\crefname{figure}{Figure}{Figures}
\setcounter{secnumdepth}{3}

\begin{center}
  {\huge\bfseries Supplementary Information}\\[1.5em]
\end{center}
 
\vspace{1em}


\section{Optimization of the centrifugation step}\label{sec:SI_centr}
Density gradient ultracentrifugation (DGU) experiments were carried out to separate thin and ultrathin BN nanotubes from impurities and larger tubes. By varying the centrifugation time from \qtyrange[range-units=single]{1}{2.75}{\hour}, we determined the minimum duration required for effective separation between these species. These tests confirmed that durations shorter than \qty{2.75}{\hour} were insufficient for the BNNTs to reach their isopycnic positions (\ie the density-matched equilibrium positions). Consequently, the BNNTs failed to resolve into discrete bands, resulting instead in a single mixed layer with low tube concentration. Analysis of these fractions revealed a broad diameter distribution (\qtyrange[range-phrase=--, range-units=single]{0.5}{8.0}{\nano\meter}), indicating a lack of selectivity compared to the equilibrium reached at \qty{2.75}{\hour}.\\

\begin{figure}[!h]
    \centering
    \includegraphics[width=0.4\linewidth]{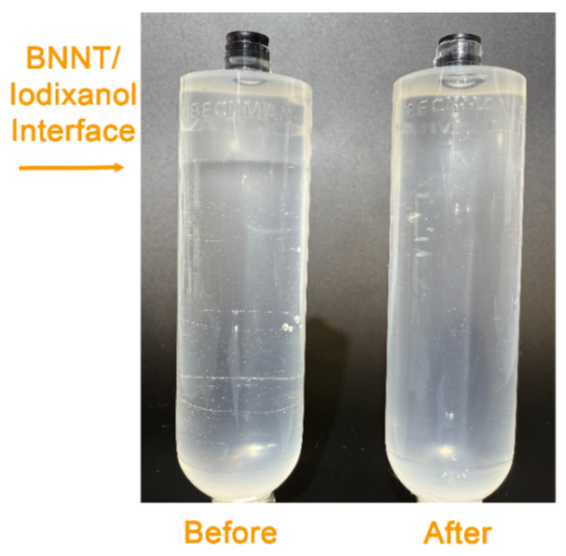}
    \caption{Centrifuge tubes with iodixanol solution and BNNT suspension before and after DGU sorting.}
    \label{fig:SI_centr_tubes}
\end{figure}

\begin{figure}[!h]
    \centering    
    \includegraphics[width=\textwidth]{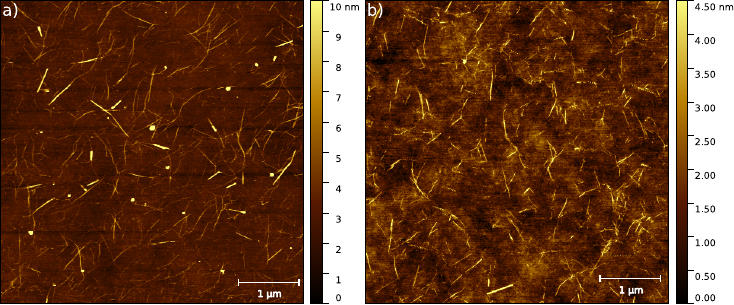}

    \caption{Representative AFM images of BNNTs from a) Band 1 and b) Band 2 deposited onto silicon (see Fig. 2b in the main text for the diameter distribution of each Band).}
    \label{fig:SI_afm}
    
    \phantomsubcaption{\label{fig:SI_afm_a}}
    \phantomsubcaption{\label{fig:SI_afm_b}}
    
\end{figure}

\begin{table}[!ht]
    \centering
    \begin{tabular}{c|c|c|c|c|c}
        fraction & A (\unit{\per\nano\meter}) & $\mu$ & $\sigma$ & E (\unit{\nano\meter}) & SD (\unit{\nano\meter}) \\
        Band~1 & \num{22 +- 2} & \num{0.21 +- 0.05} & \num{0.46 +- 0.04} & \num{1.4 +- 0.1} & \num{0.7 +- 0.1} \\
        Band~2 & \num{27 +- 2} & \num{0.88 +- 0.04} & \num{0.44 +- 0.04} & \num{2.7 +- 0.1} & \num{1.2 +- 0.1}\\
    \end{tabular}
    \caption{Lognormal fit ($Ae^{-\left(\ln{d}-\mu\right)^2 / 2\sigma^2}/\left(d\sigma\sqrt{2\pi}\right)$, where $d$ is the tube diameter) parameters based on the BNNT diameter distributions of Fig.~2b in the main text for Band~1 and Band~2. The mean value ($E = e^{\mu + \sigma^2 / 2}$) and the standard deviation ($SD = E \cdot \sqrt{e^{\sigma^2} - 1}$) are reported as well.}
    \label{tab:SI_lognorm_fit}
\end{table}

\begin{figure}[h]
    \centering
    \includegraphics[width=0.8\linewidth]{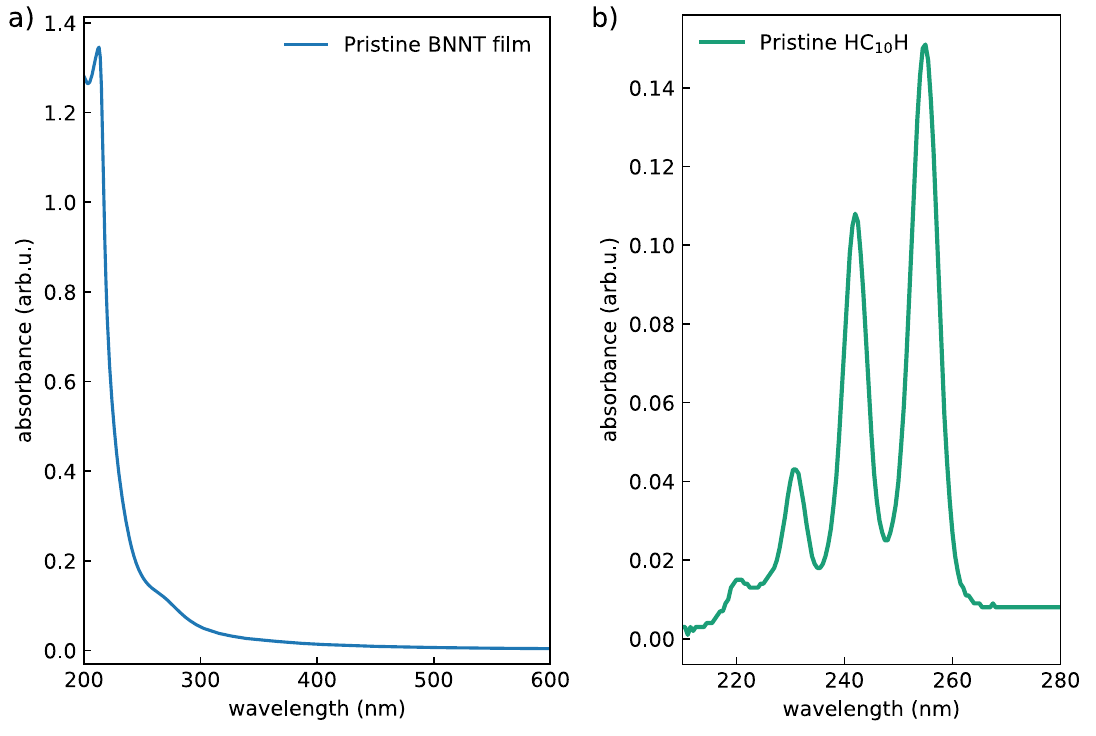}
    \caption{UV-Vis absorption spectra of a) the pristine BNNTs film and b) the pristine HC$_{10}$H solution in cyclohexane.}
    \label{fig:SI_uvvis}

    \phantomsubcaption{\label{fig:SI_uvvis_a}}
    \phantomsubcaption{\label{fig:SI_uvvis_b}}
\end{figure}

\section{Optimization of the polyyne-to-carbyne conversion process} \label{sec:SI_growth}
\begin{figure}[!h]
    \centering    
    \includegraphics[width=\textwidth]{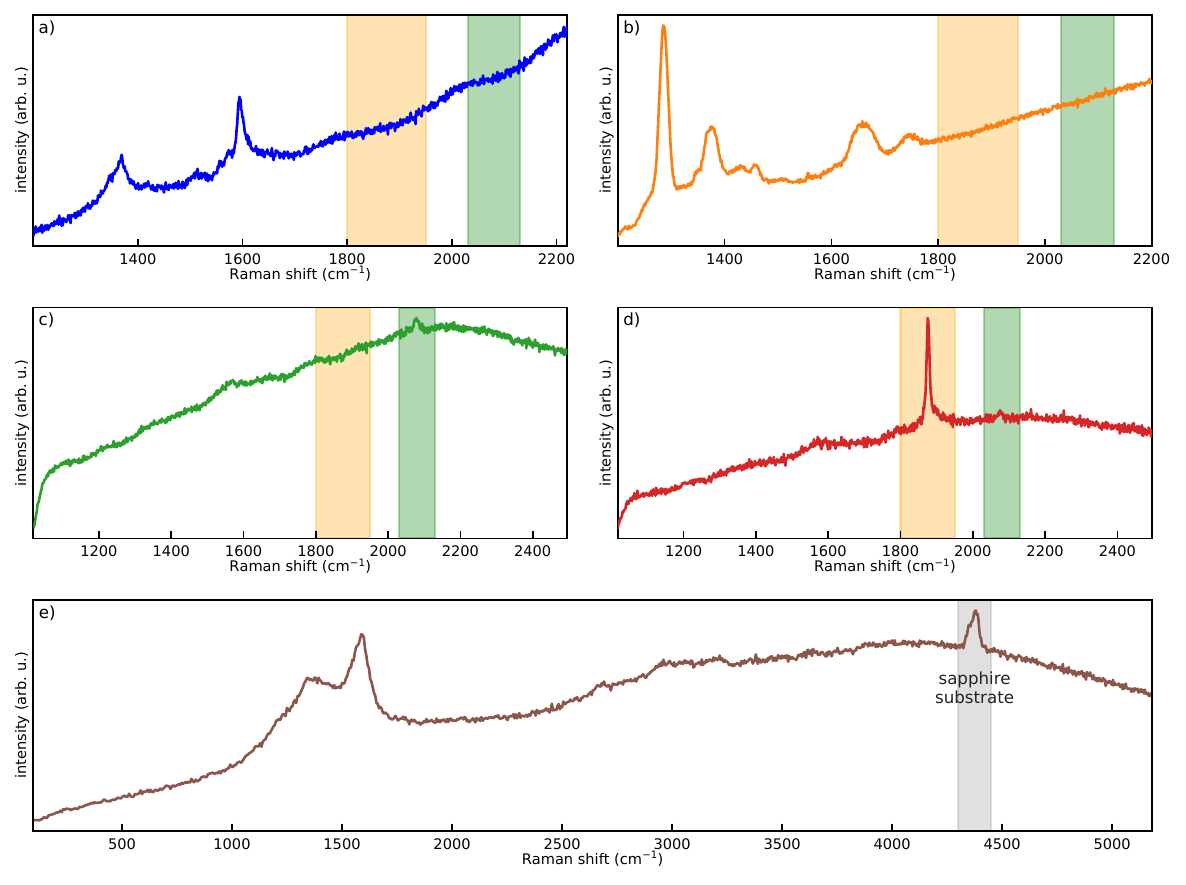}

    \caption{Raman spectra of a) the as-prepared BNNT thin film, b) the MCE filter, and after HC$_{10}$H encapsulation and annealing at c) \qty{300}{\celsius} and d) \qty{350}{\celsius}. The orange and olive green shaded areas highlight the vibrational regions of the confined carbyne C mode and HC$_{10}$H ECC mode. e) Sample position with other-than-carbyne carbon contributions. A gray area partially obscures the signal from the sapphire substrate. All Raman spectra have been collected using a \qty{532}{\nano\meter} excitation (see details in the Methods section in the main text).}
    \label{fig:SI_growth}
    
    \phantomsubcaption{\label{fig:SI_growth_a}}
    \phantomsubcaption{\label{fig:SI_growth_b}}
    \phantomsubcaption{\label{fig:SI_growth_c}}
    \phantomsubcaption{\label{fig:SI_growth_d}}
    \phantomsubcaption{\label{fig:SI_growth_e}}
    
\end{figure}

\noindent The pristine Raman spectrum of the BNNT thin film before encapsulation of HC$_{10}$H and polyyne-to-carbyne conversion is shown in \cref{fig:SI_growth_a}. The characteristic $\mathrm{E_{2g}}$ mode of BNNTs cannot be resolved under visible, non-resonant excitation (\qty{2.33}{\electronvolt})~\cite{arenal2006raman}, while the features centered at \qtylist[list-separator=and, list-units=single]{\approx 1370;1595}{\ramcm} are more likely associated with residual h-BN impurities remaining after purification~\cite{arenal2006raman}, spurious carbonaceous species already encapsulated within the nanotube cores in the as-grown parent material, or residues of the MCE filter used during the wet transfer procedure, whose Raman spectrum is reported in \cref{fig:SI_growth_b}.\\
\noindent A consideration regarding the comparison between the redshift of the longitudinal stretching mode of HC$_{10}$H upon encapsulation in CNT and BNNT is required here. This mode is redshifted by \qtyrange[range-phrase=--, range-units=single]{\approx 38}{41}{\ramcm} compared to the isolated molecule in solution (\qtyrange[range-phrase=-, range-units=single]{2120}{2123}{\ramcm})~\cite{marabotti2022electron, tabata2006raman}. The same molecule confined in CNTs is found between \qtyrange[range-phrase=~and~, range-units=single]{2058}{2074}{\ramcm}, \ie redshifted by \qtyrange[range-phrase=-, range-units=single]{\approx 49}{62}{\ramcm}~\cite{nishide2006single, nishide2007raman, tang2024encapsulation, moura2009charge, moura2011dielectric}. In HC$_{10}$H@CNT, this redshift slightly decreases with decreasing host diameter from \qty{62}{\ramcm} at \qty{1.40}{\nano\meter} to \qty{49}{\ramcm} at \qty{1.35}{\nano\meter}, even though a clear diameter-frequency relation has not been established yet for these systems, unlike in carbyne@CNT~\cite{nishide2006single}. For these reasons, a residual diameter contribution to the smaller BNNT value cannot be excluded from the present data. Thus, in the main text, we note only that \qty{38}{\ramcm} lies below the entire reported CNT range.

\noindent We performed a series of annealing experiments to optimize the conversion of encapsulated carbon atomic wires (\ie HC$_{10}$H) into carbyne chains. Four annealing temperatures were tested, \qtylist{300; 350; 400; 450}{\celsius}, each with a fixed duration of \qty{8}{\hour}.\\
\noindent Annealing at \qty{300}{\celsius} resulted in an attenuation of the HC$_{10}$H Raman feature, indicating partial degradation of the encapsulated wires rather than efficient carbyne growth, as shown in~\cref{fig:SI_growth_c}. Furthermore, the vacuum annealing efficiently removed hBN impurities from the film, as evidenced by the disappearance of the \qty{1370}{\ramcm} Raman peak. At \qty{350}{\celsius}, the spectra showed the coexistence of the HC$_{10}$H mode at \qty{2082}{\ramcm} and emerging C mode-like peaks in the \qtyrange[range-units=single, range-phrase=--]{1870}{1890}{\ramcm} range, indicating partial conversion (see~\cref{fig:SI_growth_d}).\\
\noindent Upon annealing at \qty{400}{\celsius}, the C mode became dominant across most probed positions (\cref{fig:fig2e} in the main text), although residual molecular signatures were still occasionally observed, as at other annealing temperatures~(\cref{fig:SI_growth_d}), indicating spatially heterogeneous conversion within the film.\\
\noindent Annealing above \qty{400}{\celsius} led to the complete loss of detectable sp-carbon-related Raman signals, suggesting degradation of the BNNT host and release of the encapsulated chains, followed by thermal decomposition or crosslinking.\\
\noindent These results identify \qty{400}{\celsius} for \qty{8}{\hour} as the optimal condition for achieving efficient polyyne-to-carbyne conversion while preserving the structural integrity of BNNT hosts.\\
\noindent Alongside the carbyne C mode signal, Raman spectra acquired after the optimized annealing across the sample occasionally reveal contributions from additional carbon species. These spectra are characterized by broad features from \qtyrange[range-units=single]{1000}{1600}{\ramcm}. A representative example is shown in~\cref{fig:SI_growth_e}. These broad signals can originate from amorphous carbon within or around the BN tubes, but can also indicate signatures of graphene nanoribbons (GNRs) or ribbon-like sp$^2$ carbon species encapsulated within BNNTs~\cite{barzegar2016synthesis, cadena2023encapsulation}. Considering that GNRs with different edge geometries (zigzag or armchair) and widths, including possible folding and defects, can be present within the same nanotube or bundle of BNNTs, the resulting Raman spectrum is a convolution of all single species characteristic vibrational responses. In the end, this can produce the broad Raman band observed in~\cref{fig:SI_growth_e}~\cite{talyzin2011synthesis, verzhbitskiy2016raman, cai2010atomically, llinas2017short, cataldo2010graphene, gillen2010raman, yang2011observation}. Nevertheless, thermal conversion of sp-carbon chains encapsulated in BNNTs into GNRs has never been reported to date. BNNT-encapsulated aromatic precursors, instead, have been used for this purpose at comparable annealing temperatures (\qty{430}{\celsius})~\cite{barzegar2016synthesis, cadena2023encapsulation}.\\
\noindent The origin of these amorphous or sp$^2$ species is most plausibly attributed to the decomposition of the HC$_{10}$H precursor outside or within  BNNTs. While it is known that H-capped polyynes degrade instantly when exposed to air, due to crosslinking and oxidation reactions, these wires remain thermally stable up to \qty{350}{\celsius} when confined in carbon nanotubes with a diameter of \qty{1.4 \pm 0.1}{\nano\meter}~\cite{nishide2006single}. The authors of this study suggested that when the cavity is too wide to force linear chain formation, the annealed short wires can transform into planar or ribbon-like carbon structures. This behavior is consistent with prior observation from Schuster~\etal~in CNT-based systems, where carbon precursor decomposition in large diameter (\qty{> 1.35}{\nano\meter}) tubes yields the growth of another CNT layer(s) rather than carbyne~\cite{schuster2025quantifying}. Based on these results, we might consider the loss of sp-character in some of the BNNTs without, however, having direct evidence of the specific diameter range that leads to this phenomenon.

\clearpage
\section{Laser-induced photodegradation and correction of polarization-dependent Raman measurements} \label{sec:SI_degr}
\begin{figure}[!h]
    \centering
    \includegraphics[width=0.65\linewidth]{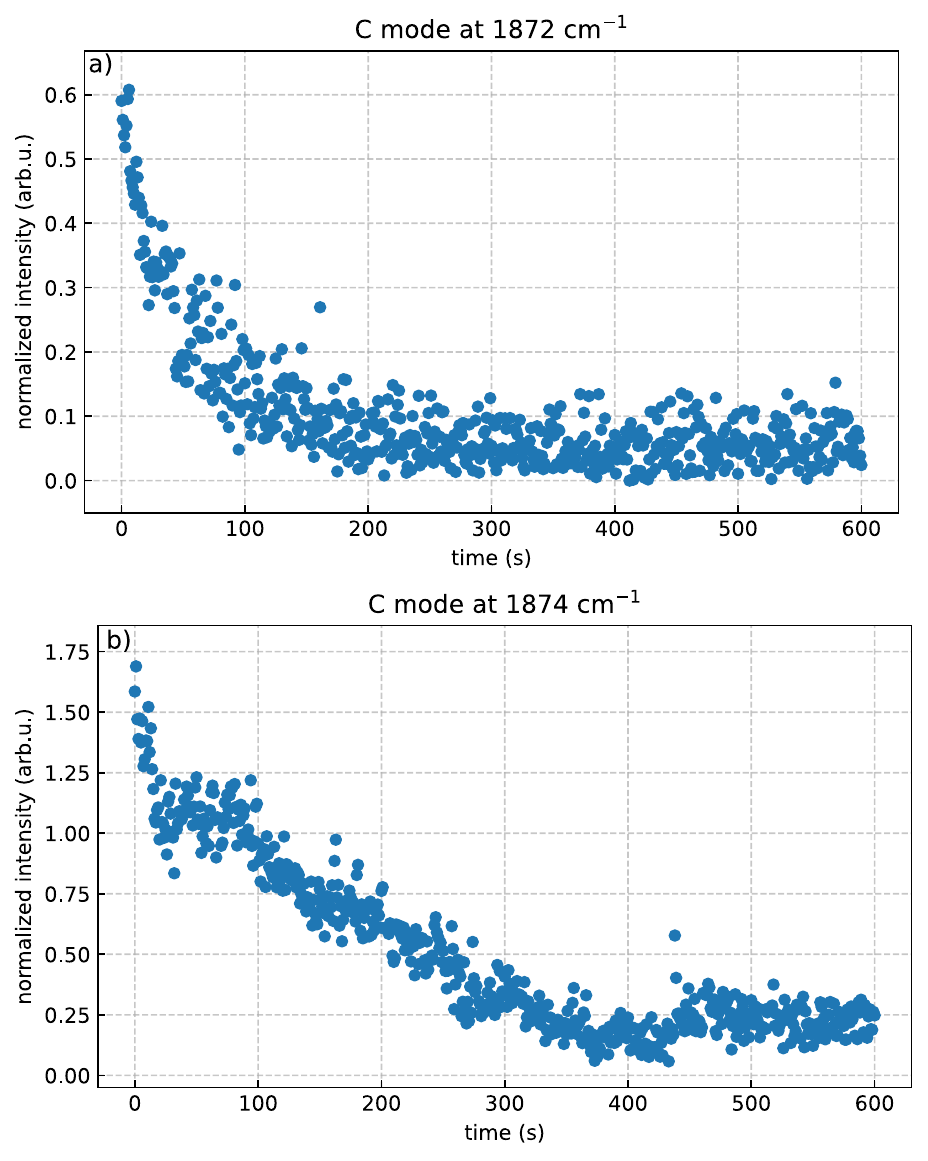}
    \caption{Laser-induced degradation of two different near-ideal carbyne chains under continuous irradiation under fixed excitation wavelength (\qty{532}{\nano\meter}), acquisition time (\qty{1}{\second}), laser power (\qty{0.16}{\milli\watt}), and light polarization (circular). For each spectrum, the area of the fitted C mode peak is normalized using the area of the sapphire (substrate) signal at \qty{\approx 4350}{\ramcm}.}
    \label{fig:SI9}

    \phantomsubcaption{\label{fig:SI9a}}
    \phantomsubcaption{\label{fig:SI9b}}
\end{figure}

The photostability of carbyne@BNNTs was evaluated by monitoring the C mode intensity as a function of continuous irradiation time at fixed excitation (\qty{532}{\nano\meter}) conditions (see \cref{fig:SI9}). A gradual and monotonic decrease of the C mode integrated area is observed upon prolonged exposure, exhibiting different timescales depending on the specific chain and environment (\cref{fig:SI9a,fig:SI9b}). By normalizing the C mode area to that of the substrate (sapphire) in each spectrum, we exclude any defocusing process as the origin of the intensity loss. The observed decay reflects intrinsic laser-induced degradation phenomena rather than optical misalignment.\\
The possibility that the Raman signal loss originates from damage or destruction of the BNNT host, leading to exposure to air and subsequent degradation of the encapsulated chains, can be safely excluded under our experimental conditions (\qty{0.16}{\milli\watt} excitation power, circular polarization, \qty{532}{\nano\meter}). Boron nitride nanotubes are known for their exceptional chemical and thermal stability, as well as their large bandgap, which strongly limits absorption at the excitation wavelength. Under such irradiation conditions, no structural modification of the BNNT is expected. We therefore attribute the observed decay to degradation processes affecting the carbyne chains themselves. A detailed investigation of the underlying degradation mechanisms is beyond the scope of the present work.

\vspace{5em}
\begin{figure}[!h]
    \centering    
    \includegraphics[width=0.9\textwidth]{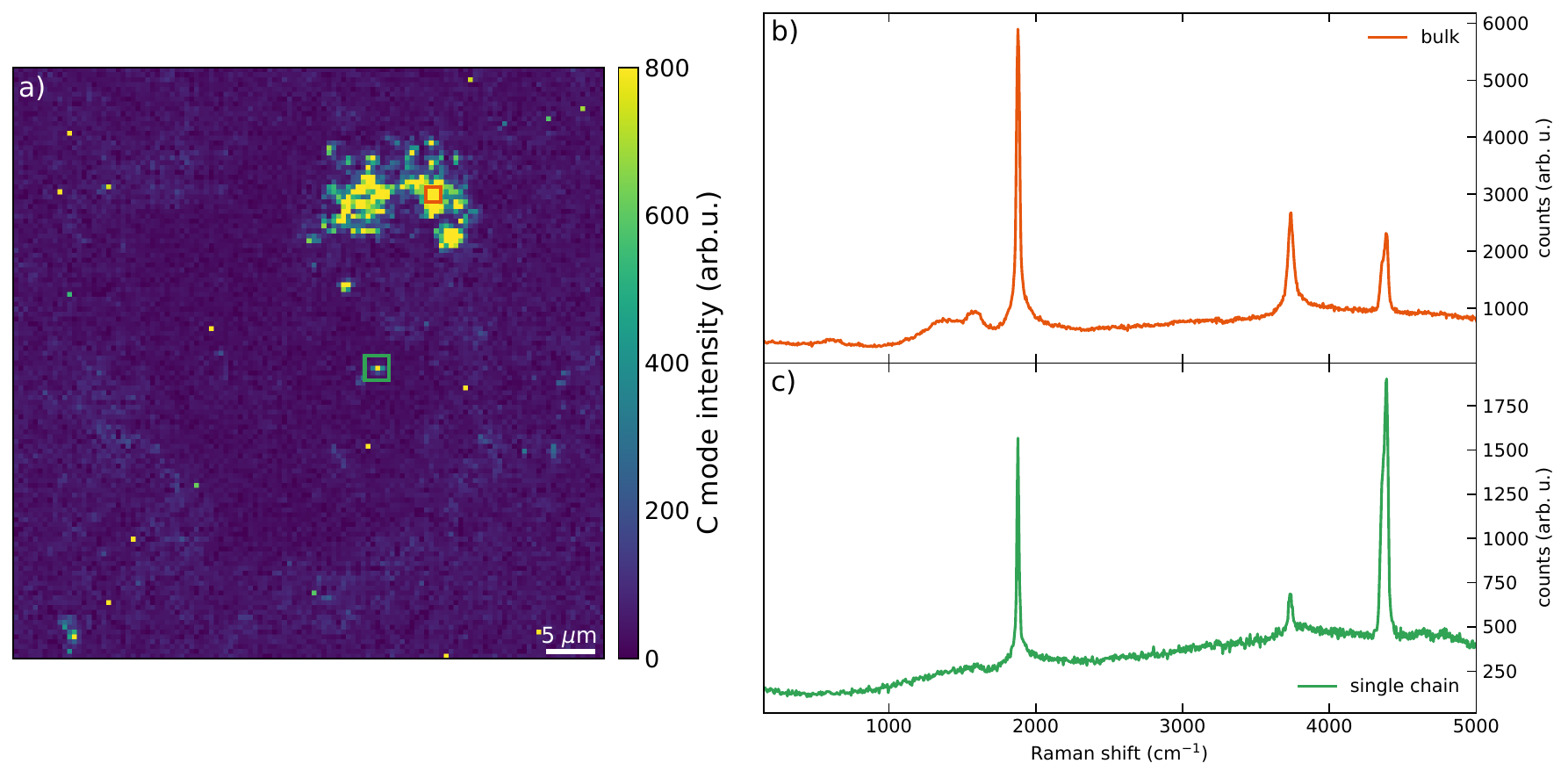}

    \caption{a) Confocal Raman map showing the C mode intensity in a $60 \times 60$ \unit{\micro\meter} region of the sample, acquired with a \qty{532}{\nano\meter} laser and a step size of \qty{500}{\nano\meter}. Raman spectra extracted from the map correspond to b) a bulk region (red square) and c) a single chain (green square).}
    \label{fig:SI_bigmap}
    
    \phantomsubcaption{\label{fig:SI_bigmap_a}}
    \phantomsubcaption{\label{fig:SI_bigmap_b}}
    \phantomsubcaption{\label{fig:SI_bigmap_c}}
    
\end{figure}

\begin{figure}[!h]
    \centering    
    \includegraphics[width=0.8\textwidth]{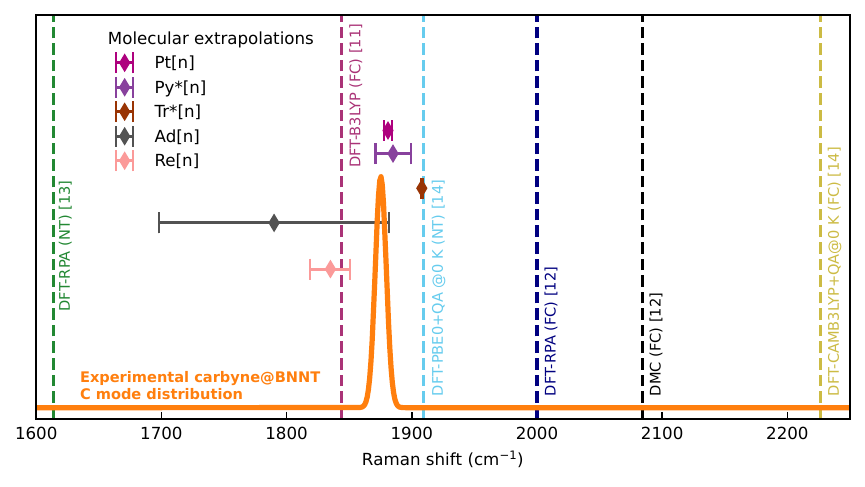}

    \caption{Comparison between the distribution of measured carbyne@BNNT C modes across 245 locations (orange thick line,~\cref{fig:fig4a} in the main text), theoretical values for both the free-chain (FC) and a chain inside a dielectric environment (NT)~\cite{milani2008first, mostaani2016quasiparticle, ramberger2021new, romanin2021dominant}, and extrapolations from molecular (CAWs) series with different terminations~\cite{gao2022advances, arora2023monodisperse}. The theoretical values are reported also in \cref{tab:SI_theory}.}
    \label{fig:SI_distr_theo_extrap}
    
\end{figure}

\clearpage

\section{Sensitivity of the vibrational anharmonicity comparison} \label{sec:SI_anh}

\noindent The four carbyne@BNNT chains for which overtones could be resolved span \qty{8.2}{\ramcm} in the fundamental frequency (see~\cref{tab:SI_anh}). Over such a narrow interval the universal trend predicts a variation of the total anharmonic redshift of \qty{0.2}{\ramcm} (2C) and \qty{0.9}{\ramcm} (3C), far below the measurement scatter. In this framework, the statistical quantity to be compared with the trend is their mean.

\noindent The fitting uncertainties reported in~\cref{tab:SI_anh} are not the relevant scale for this comparison, since they reflect the precision of the individual Lorentzian fits alone. The trend itself is defined by data scattering with a residual dispersion of \qty{2.6}{\ramcm} (2C) and \qty{8.6}{\ramcm} (3C). The four carbyne@BNNT chains scatter by \qty{3.0}{\ramcm} and \qty{6.5}{\ramcm} respectively, \ie by the same amount as the systems that define the trend.

\begin{figure}[!h]
    \centering
    \includegraphics[width=\linewidth]{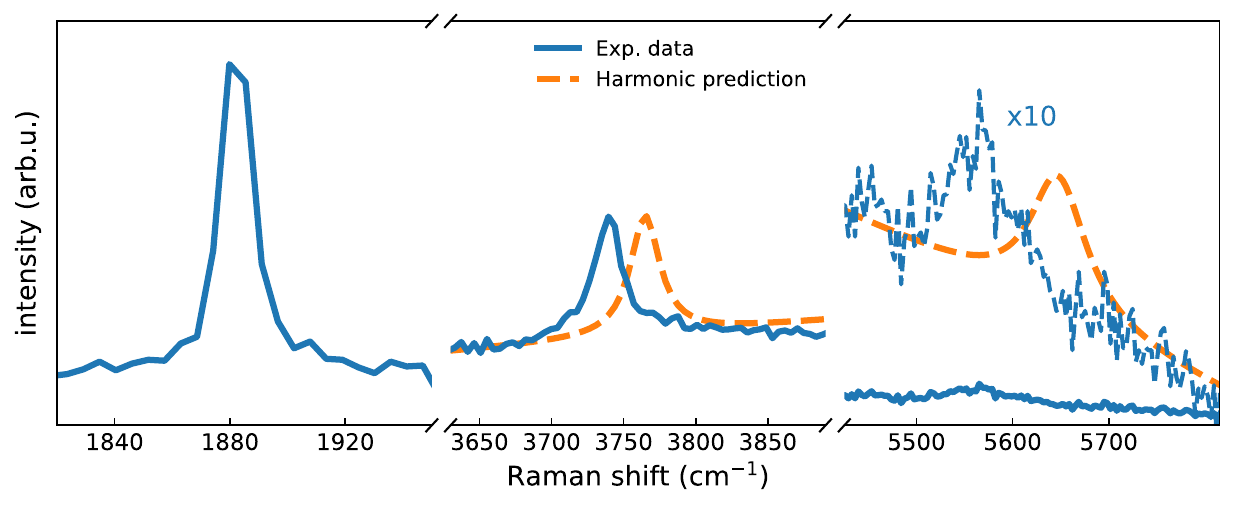}
    \caption{Experimental Raman spectrum of a near-ideal carbyne chain with $\omega_C = 1882\ cm^{-1}$ compared with a hypothetical spectrum based on a harmonic vibrational potential.}
    \label{fig:SI_overtones}
\end{figure}

\begin{table}[!h]
    \centering
    \begin{tabular}{c|c|c|c|c}
        C mode (\unit{\ramcm}) & 2C mode (\unit{\ramcm}) & 3C mode (\unit{\ramcm}) & $\tilde{\nu}_{tot,2}$ & $\tilde{\nu}_{tot,3}$ \\
        \num{1882.4 +- 0.1} & \num{3739.1 +- 0.3} & \num{5566 +- 2} & \num{25.7 +- 0.3} & \num{81 +- 2} \\
        \num{1878.1 +- 0.3} & \num{3737.2 +- 0.3} & \num{5566 +- 2} & \num{19.0 +- 0.7} & \num{68 +- 2} \\
        \num{1884.2 +- 0.1} & \num{3745.0 +- 0.6} & \num{5573 +- 5} & \num{23.4 +- 0.6} & \num{80 +- 5} \\
        \num{1876 +- 1} & \num{3731 +- 1} & \num{5555 +- 2} & \num{21 +- 2} & \num{73 +- 4} \\
    \end{tabular}
    \caption{Fundamental C mode and overtone (2C and 3C) frequencies of carbyne@BNNT chains, together with calculated total anharmonic redshifts for the 2C ($\Delta\tilde{\nu}_{tot,2} = 2\cdot\tilde{\nu}_1 - \tilde{\nu}_2$) and 3C ($\Delta\tilde{\nu}_{tot,3} = 3\cdot\tilde{\nu}_1 - \tilde{\nu}_3$) modes. Fitting errors of experimental Raman peaks and calculated errors of total anharmonic redshifts are reported as well.}
    \label{tab:SI_anh}
\end{table}

\section{Extraction of the bond length alternation from the C mode frequency}
\label{sec:SI_bla}

\subsection{H\"{u}ckel model chain}

The BLA reported in the main text is obtained from the measured C mode frequency through three steps. The C mode frequency fixes the dimerization parameter ($\bar{b}$) of the H\"{u}ckel model of Ref.~\cite{milani2009connection}. The optical gap is then obtained from the same C mode frequency through the empirical correlation established from carbyne@CNT~\cite{martinati2022electronic}. These two values together set the absolute length scale of the alternation. No quantity other than $\omega_C$ is measured, so the BLA is a structural re-parametrization of the vibrational observable rather than an independent determination.

\paragraph{Step 1: $\omega \rightarrow \bar{b}$.} The model describes the $\pi$-electron system at the H\"{u}ckel level and the $\sigma$ system as a harmonic spring of force constant $k_\sigma$. Defining $\bar{b}$ as the dimensionless dimerization parameter, the two lattice sums are

\begin{equation}
F(\bar{b}) = \int_{-\pi}^{\pi} \frac{1-\cos \theta}
{\left[1+\cos \theta + \bar{b}^{\,2}(1-\cos \theta)\right]^{1/2}}\,\mathrm{d}\theta ,
\qquad
C(\bar{b}) = \int_{-\pi}^{\pi} \frac{\sin^{2} \theta}
{\left[1+\cos \theta + \bar{b}^{\,2}(1-\cos \theta)\right]^{3/2}}\,\mathrm{d}\theta ,
\end{equation}

\noindent and the frequency of the longitudinal optical mode at $\Gamma$, \ie the C mode,
follows from

\begin{equation}
F_R = k_\sigma\left[1-\frac{C(\bar{b})}{F(\bar{b})}\right] ,
\qquad
\omega_R^{2} = \frac{4F_R}{m_C} .
\label{eq:si_omega}
\end{equation}

\noindent $\omega_R(\bar{b})$ is monotonically increasing over the whole domain, so the inversion $\bar{b}(\omega)$ is unique and it is performed here numerically. This step depends on $k_\sigma$ alone and is independent of the optical gap. We adopt the same value of $k_\sigma$ used in Ref.~\cite{milani2009connection}, \ie $k_\sigma = \qty{16.75}{\milli\dyne / \angstrom}$.

\paragraph{Step 2: $\omega \rightarrow E_g$.} We use the linear correlation between C mode frequency and optical gap reported for carbyne@CNT~\cite{martinati2022electronic}, $E_g = A\cdot\omega + B$, with $A = \qty{6.17(62)e-3}{\electronvolt}\text{/}\unit{\ramcm}$ and $B = \qty{-9.22(113)}{\electronvolt}$. The correlation is calibrated on carbyne@CNT C mode components spanning \qtyrange[range-units=single, range-phrase=--]{1790}{1857}{\ramcm}.

\paragraph{Step 3: $(\bar{b}, E_g) \rightarrow \Delta r$.} The gap sets the energy scale $\beta_0 = E_g/4\bar{b}$, from which
$\beta' = (\pi^{1/2}/2^{1/4})\sqrt{k_\sigma\beta_0/F(\bar{b})}$ and

\begin{equation}
\Delta r = \frac{E_g}{2\beta'}
= \frac{2^{1/4}}{\sqrt{\pi}}\sqrt{\frac{F(\bar{b})\,\bar{b}\,E_g}{k_\sigma}} .
\label{eq:si_bla}
\end{equation}

\noindent At fixed $\omega$ the dimerization parameter $\bar{b}$ is fixed, so $\Delta r \propto \sqrt{E_g}$.

\subsection{Numerical results}

Table~\ref{tab:si_bla} lists the extracted BLA values ($\Delta r$) and the model quantities across the frequency range relevant to this work. For carbyne@BNNT, $\langle\omega_C\rangle_{\mathrm{BNNT}} = \qty{1875.23}{\ramcm}$ gives $\bar{b} = 0.1113$, $F_R =
\qty{6.221}{\milli\dyne / \angstrom}$, $E_g = \qty{2.350}{\electronvolt}$, $\beta_0 = \qty{5.277}{\electronvolt}$, $\beta' =
\qty{9.124}{\electronvolt\per\angstrom}$ and $\Delta r = \qty{0.1288}{\angstrom}$. The FWHM of the measured distribution (\qty{10.31}{\ramcm}) maps onto a BLA interval of \qty{3.1}{\milli\angstrom}, whereas the carbyne@CNT frequency range \qtyrange{1760}{1870}{\ramcm} maps onto \qtyrange{0.0941}{0.1272}{\angstrom}, a span of \qty{33.1}{\milli\angstrom} corresponding to a variation of \qty{25}{\percent}.

\begin{table}[h]
\centering
\begin{tabular}{ccccc}
\hline
$\omega$ (\unit{\ramcm}) & $\bar{b}$ & $C/F$ & $E_g$ (\unit{\electronvolt}) & $\Delta r$ (\unit{\angstrom}) \\
\hline
1760.00 & 0.0739 & 0.6728 & 1.639 & 0.0941 \\
1800.00 & 0.0858 & 0.6578 & 1.886 & 0.1061 \\
1840.00 & 0.0988 & 0.6424 & 2.133 & 0.1181 \\
1870.00 & 0.1094 & 0.6307 & 2.318 & 0.1272 \\
\textbf{1875.23} & \textbf{0.1113} & \textbf{0.6286} & \textbf{2.350} & \textbf{0.1288} \\
1880.00 & 0.1131 & 0.6267 & 2.380 & 0.1302 \\
1890.00 & 0.1169 & 0.6227 & 2.441 & 0.1333 \\
1900.00 & 0.1208 & 0.6187 & 2.503 & 0.1363 \\
\hline
\end{tabular}
\caption{Model quantities as a function of the C mode frequency, for $k_\sigma = \qty{16.75}{\milli\dyne / \angstrom}$. The row in bold corresponds to $\langle\omega_C\rangle_{\mathrm{BNNT}}$. Digits reflect the numerical precision of the model inversion, not the accuracy of $\Delta r$, which is set by the approximations and the sensitivity discussed in Section~S.5.3.}
\label{tab:si_bla}
\end{table}

\subsection{Model approximations, uncertainty, and sensitivity} \label{sec:SI_bla_approx}

The model is used here as published in Ref.~\cite{milani2009connection} and the approximations are listed in the following. The $\pi$-electron system is treated at the H\"{u}ckel level in a mean-field approximation, and correlation effects may become relevant in the strong-distortion limit. The $\sigma$ system is described as a classical harmonic spring, so the model contains no quantum nuclear fluctuations. The force constant $k_\sigma$ is transferred from small hydrocarbons and is not fitted to the system under study. The gap correlation of Ref.~\cite{martinati2022electronic} is calibrated on carbyne@CNT and is applied here \qty{18}{\ramcm} beyond the upper end of its calibration range. It is worth noticing that the gap entering Eq.~\eqref{eq:si_bla} is the single-particle $\pi$--$\pi^*$ gap of the H\"{u}ckel model, while Ref.~\cite{martinati2022electronic} provides the optical gap, so the exciton binding energy is implicitly neglected. This approximation is common to every system analyzed with the same model chain and therefore does not invalidate the comparisons drawn in the main text.

\noindent The dimerization parameter ($\bar{b}$) is not affected by any of the approximations entering the energy scale, since it follows from $\omega$ and $k_\sigma$ alone. For carbyne@BNNT, $\bar{b} = 0.1113$  against $\bar{b} = 0.0739$--$0.1094$ over the carbyne@CNT range (Table~\ref{tab:si_bla}). The ordering of the two systems in the dimerization parameter is therefore independent of the gap correlation, which enters only in converting $\bar{b}$ into an absolute length.

\noindent The parameters ($A$ and $B$) of the optical gap correlation are strongly anticorrelated, as in any linear regression, but their covariance is not reported in Ref.~\cite{martinati2022electronic}. Propagating the two quoted uncertainties as if independent gives $\sigma(E_g) \approx \qty{1.6}{\electronvolt}$ at $\langle\omega_C\rangle_{\mathrm{BNNT}}$, larger than the gap itself and evidently meaningless for a correlation that reproduces its calibration set to within a few tens of \unit{\milli\electronvolt}. We therefore quote no propagated uncertainty on $\Delta r$, and characterize instead its sensitivity to the gap. Since $\bar{b}$ is fixed by $\omega$ alone, $\Delta r \propto \sqrt{E_g}$ and the dependence is damped. Therefore, varying $E_g$ by $\pm$\qty{0.15}{\electronvolt} about \qty{2.350}{\electronvolt} moves $\Delta r$ from \qty{0.1246}{\angstrom} to \qty{0.1329}{\angstrom}, \ie by \qty{4}{\milli\angstrom}. This is smaller than the spread among the first-principles values it is compared with in Table~\ref{tab:SI_theory}, and an order of magnitude smaller than the host-induced modulation quantified in the main text.

\noindent The residual bias has a known sign. The correlation is calibrated on carbyne@CNT, where the chain is dielectrically screened by the host. Screening lowers the optical gap at fixed alternation, so applying the same correlation to a less screening host (\ie BNNT) underestimates $E_g$ and, through $\Delta r \propto \sqrt{E_g}$, underestimates the BLA. The value reported for carbyne@BNNT is therefore a lower bound, and the free-chain calculations it is compared with in the main text lie at or above it. The differential comparison between the two systems is in any case unaffected: both are mapped through the same correlation and the same model chain, so a systematic error in $A$, $B$ or $k_\sigma$ displaces both in the same direction.

\subsection{Comparison with first-principles calculations}

Table~\ref{tab:SI_theory} collects published BLA and longitudinal optical phonon frequencies for the infinite chain, separated according to whether the calculation describes a free chain or a chain in a screening environment. The BLA extracted for carbyne@BNNT, \qty{0.1288}{\angstrom}, falls at the lower edge of the free chain values obtained by methods that treat electron correlation explicitly (RPA, DMC, quantum-anharmonic DFT), which agree among themselves within \qty{7}{\milli\angstrom}, while the values extracted for carbyne@CNT fall among those computed with screening. The DFT-B3LYP free-chain value, \qty{0.0844}{\angstrom}, lies below the entire screened set, confirming that the spread among calculations is dominated by the level of theory rather than by the environment, and a structural comparison is meaningful only among methods that recover correlation. Considering the C mode frequencies, these correlated free-chain calculations place the longitudinal optical mode between \qty{2000}{\ramcm} and \qty{2227}{\ramcm}, well above $\langle\omega_C\rangle_{\mathrm{BNNT}}$, while the nanotube DFT-RPA (random phase approximation) calculation places it at \qty{1614}{\ramcm}, well below the carbyne@CNT range. No clear BLA--frequency relation is observed across the calculations themselves, \textit{e.g.}, quantum Monte Carlo and quantum-anharmonic calculations provide BLA values within \qty{2}{\milli\angstrom} of one another but frequencies differing by \qty{140}{\ramcm}.\\

\noindent Such a discrepancy between calculated BLA and C mode frequency values is expected and justifies the comparison drawn in the main text between BLA values rather than between frequencies. The computed phonon frequency of a $\pi$-electron conjugated chain is far more sensitive to the level of theory than the computed structure. For instance, within Ref.~\cite{romanin2021dominant} alone, including quantum anharmonicity and relaxing the lattice accordingly changes the BLA from \qty{0.0973}{\angstrom} to \qty{0.0980}{\angstrom} (\qty{0.7}{\percent}), while the same treatment moves the frequency from \qty{1986.6}{\ramcm} to \qty{1880.7}{\ramcm} (\qty{5.3}{\percent}). By comparison, our BLA value carries the limitation of a classical effective BLA obtained from a measured frequency under the approximations listed above. Agreement in the structural parameter alongside disagreement in the frequency is expected, and the structural comparison is the one that both sides support. For the same reason, the proximity of $\langle\omega_C\rangle_{\mathrm{BNNT}}$ to the PBE0 frequency does not assign carbyne@BNNT to a screened environment. The two values differ by \qty{31}{\milli\angstrom} in the structural parameter, while their frequencies agree only because the frequency is the quantity that shifts by \qty{5.3}{\percent} under a change of treatment.

\begin{table}[h]
\centering
\begin{tabular}{lccl}
\hline
method & BLA (\unit{\angstrom}) & $\omega_{LO}$ (\unit{\ramcm}) & ref. \\
\hline
\multicolumn{4}{l}{\textit{free chain}} \\
DFT-B3LYP                           & 0.0844 & 1844 & \cite{milani2008first} \\
diffusion quantum Monte Carlo (DMC) & 0.136 & 2084 & \cite{mostaani2016quasiparticle} \\
DFT-RPA                 & 0.129 & 2000 & \cite{ramberger2021new} \\
DFT-CAMB3LYP + quantum anharmonicity @\qty{0}{\kelvin} & 0.1348 & 2226.5 & \cite{romanin2021dominant} \\
\textbf{carbyne@BNNT} & \textbf{0.1288} & \textbf{1875.2} & \textbf{this work} \\
\hline
\multicolumn{4}{l}{\textit{chain in a screening environment}} \\
DFT-RPA (CNT)                & 0.091 & 1614 & \cite{ramberger2021new} \\
DFT-PBE0 + quantum anharmonicity @\qty{0}{\kelvin}  & 0.0980 & 1880.7 & \cite{romanin2021dominant} \\
\textbf{carbyne@CNT} & \textbf{0.094--0.127} & \textbf{1760--1870} & \textbf{this work} \\
\hline
\end{tabular}
\caption{Published BLA and longitudinal optical frequencies for the infinite chain, compared with the values extracted in this work. The PBE0 entry uses the quantum-anharmonically relaxed lattice parameter reported in the Supplementary Information of Ref.~\cite{romanin2021dominant}.}
\label{tab:SI_theory}
\end{table}

\section{Photoluminescence in carbyne@BNNT}
\label{sec:pl}

\begin{figure}[!ht]
    \centering
    \includegraphics[width=0.8\linewidth]{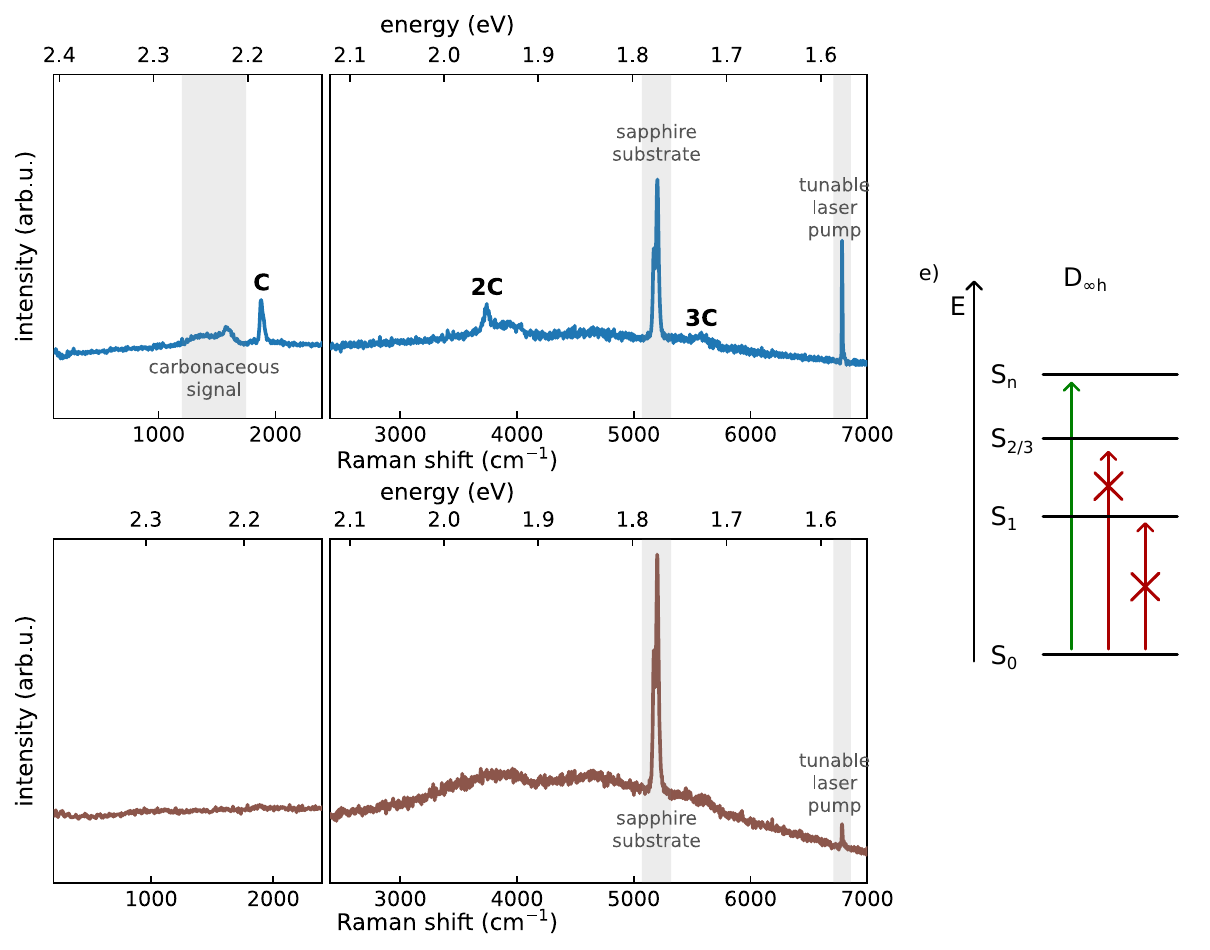}
    \caption{Raman/photoluminescence spectra of carbyne@BNNT (panels a,b) and of the sapphire substrate (panels c,d) under \qty{510}{\nano\meter} excitation. Panels a) and c) show the high-energy spectral window, panels b) and d) the low-energy one. Together they span \qty{\approx 2.42}{\electronvolt} to \qty{\approx 1.60}{\electronvolt}. e) Electronic transitions for the $\mathrm{D_{\infty h}}$ point group. Green and red arrows denote dipole-allowed and dipole-forbidden transitions respectively.}
    \label{fig:pl_spectra}

    \phantomsubcaption{\label{fig:pl_spectra_a}}
    \phantomsubcaption{\label{fig:pl_spectra_b}}
    \phantomsubcaption{\label{fig:pl_spectra_c}}
    \phantomsubcaption{\label{fig:pl_spectra_d}}
    \phantomsubcaption{\label{fig:pl_spectra_e}}
\end{figure}

\noindent To collect the PL spectrum of carbyne@BNNT, two spectral windows were acquired to cover the full emission range (see~\cref{fig:pl_spectra}). The high-energy window used a tunable bandpass filter, with a cut-on at \qty{\approx 150}{\ramcm} from the laser line (\cref{fig:pl_spectra_a,fig:pl_spectra_c}). The low-energy window used a \qty{550}{\nano\meter} long-pass filter (\cref{fig:pl_spectra_b,fig:pl_spectra_d}). Excitation power was held at \qty{\approx 50}{\micro\watt}, with acquisition times of \qty{30}{\second} and \qty{60}{\second} for each energy window, respectively. The two windows of the carbyne spectrum were joined by imposing equal first-order C mode intensities. Bare-substrate spectra were acquired under identical conditions and scaled linearly to match the signal-free regions of the corresponding window, with factors of \num{0.996} and \num{0.945}.

\noindent Besides the \num{0}--\num{0} transition at the optical gap, emission would appear in vibronic replicas one and two C mode quanta below, at \qty{2.16}{\electronvolt} and \qty{1.92}{\electronvolt}, \ie at Raman shifts of \qty{2227}{\ramcm} and \qty{4108}{\ramcm}. None of them are observed in~\cref{fig:pl_spectra_a,fig:pl_spectra_c}. 

\end{document}